\def\useieeelayout{0}
\def\showall{0}

\newcommand{\inConf}[1]{\if\showall1{\color{green!50!black}In Journal: #1}\else\if\useieeelayout1{#1}\fi\fi}

\newcommand{\inArxiv}[1]{\if\showall1{\color{blue}In ArXiV: #1}\else\if\useieeelayout0{#1}\fi\fi}

\if\useieeelayout1
\documentclass[letterpaper, 10 pt, conference]{ieeeconf}  
\IEEEoverridecommandlockouts                              %
\usepackage{longtable}
\usepackage{graphicx}
\usepackage{amsmath}
\usepackage{booktabs}
\usepackage{optidef}
 
\usepackage{amsthm}
\usepackage{amssymb}
\usepackage{amsfonts}    
\usepackage{tikz}
\usepackage{cuted}
\usepackage[dvipsnames]{xcolor}
\usepackage{makecell}
\usetikzlibrary{positioning, arrows.meta, shapes.multipart, decorations.pathreplacing}
\usepackage[american, cuteinductors]{circuitikz}
\usepackage{multicol}
\usepackage{cuted}
\usepackage{multirow}
\usepackage{caption, subcaption}
\newtheorem{theorem}{Theorem}
\newtheorem{lemma}[theorem]{Lemma}
\newtheorem{proposition}[theorem]{Proposition}
\newtheorem{definition}{Definition}
\newtheorem{problem}{Problem}
\newtheorem{assumption}{Assumption}
\newtheorem{remark}{Remark}
\newtheorem{question}{Question}
\newtheorem{conjecture}[theorem]{Conjecture}
\newtheorem{corollary}{Corollary}[theorem]
\usepackage{xcolor}
\usepackage{cite}
\usepackage[hidelinks]{hyperref}
\usepackage{comment}
\usepackage{algorithm}
\usepackage{algorithmic}
\usepackage{scalerel}
\usepackage{tabularx}
\usepackage{bm}
\usepackage[intoc, english]{nomencl}
\makenomenclature
\definecolor{steelblue}{RGB}{70,130,180}

\else
\documentclass[10pt]{article}
\usepackage{longtable}
\usepackage{graphicx}
\usepackage{amsmath}
\usepackage{optidef}
 
\usepackage{amsthm}
\usepackage{amssymb}
\usepackage{amsfonts}  
\usepackage{tikz}
\usetikzlibrary{positioning, arrows.meta, shapes.multipart, decorations.pathreplacing}
\usepackage[american, cuteinductors]{circuitikz}
\usepackage{multicol}
\usepackage{cuted}
\usepackage{multirow}
\usepackage{caption, subcaption}

\usepackage{xcolor}
\usepackage{cite}
\usepackage[hidelinks]{hyperref}
\usepackage{comment}
\usepackage{algorithm}
\usepackage{algorithmic}
\usepackage{scalerel}
\usepackage{bm}
\usepackage[intoc, english]{nomencl}
\makenomenclature
\usepackage{url}
\definecolor{steelblue}{RGB}{70,130,180}

\usepackage[preprint]{tmlr}

\title{
Unlocking the Potential of Floating Solar Photovoltaics in South America: Integrated Assessment of Energy Access, Water Security, and Grid Flexibility
}

\author{\name Soham Ghosh \email sghosh27@ieee.org \\
      \addr Black \& Veatch, Overland Park, KS 66211, US \\
\AND
\name Anik Goswami \email anik91\_go@rediffmail.com  \\
     \addr School of Electrical Engineering, Vellore Institute of Technology - Chennai, Tamil Nadu 600127, India \\ 
      \AND
      \name Krishna Kumba \email krishna.kumba@vit.ac.in \\
      \addr School of Electrical Engineering, Vellore Institute of Technology - Chennai, Tamil Nadu 600127, India 
      \AND
      \name Nabil Mohammed \email nabil.mohammed@federation.edu.au \\
      \addr Center for New Energy Transition Research (CfNETR) at Federation University, Ballarat VIC 3353 
      }

\def\month{MM}  
\def\year{YYYY} 
\def\openreview{\url{https://openreview.net/forum?id=XXXX}} 

\fi

\begin{document}

\maketitle\thispagestyle{empty}
\pagestyle{empty}

\begin{abstract}

Floating solar photovoltaic (FSPV) systems provide a promising solution for expanding clean
electricity access in regions where limited land availability constrains conventional ground-mounted photovoltaic (PV) deployment. By utilizing existing water bodies, FSPV mitigates
land-use conflicts while enhancing water and energy resource utilization. South America, with an estimated 38.26 TWh of production capacity per million acres of available water surface area, has one of the highest potentials for FSPV globally. Yet deployment remains limited compared with other continents, such as Asia. This paper presents a comprehensive integrated technical, economic, and social assessment framework to evaluate the role of FSPV in enhancing energy access, water security, and
power system flexibility in Nicaragua, Honduras, and Guyana. The assessment incorporates realistic surface coverage ratios, grid-connected plant configurations,
and utility-scale collector–substation topologies. Representative FSPV plants sized between
50 MW and 398 MW are shown to achieve annual specific energy yields of 1,500–2,000
kWh/kW and DC capacity factors exceeding 20\%, driven by favorable solar resources and
water-induced module cooling. As shall be seen in Guyana, co-locating FSPV with AI data centers can enable data center developers to generate revenue while leveraging renewable energy in their portfolios. At the El Cajón Reservoir in Honduras, FSPV deployment can reduce emissions by approximately 130.2 tCO2, 2.2 tNO2, and 2.6 tSO2 per megawatt relative to fossil-fuel-based generation. The results demonstrate competitive levelized costs of electricity through avoided land acquisition, shared hydropower infrastructure, and water conservation benefits. The proposed framework provides a scalable pathway for accelerating
FSPV deployment and strengthening the water–energy nexus across South America.

\end{abstract}


\newpage
\section*{Highlights}
\begin{itemize}
  \item South America has the highest generation potential, normalized to the available surface area of the water body.
  \item Highly favorable energy yield for FSPVs at major waterbodies in Nicaragua, Honduras, and Guyana ranges at around 2000 kWh/kW.
  \item Special use cases of colocation with AI data center plants and colocation within hydro reservoirs are discussed.
  \item  398 MWac floating solar PV use case in El Cajon dam illustrates 130.2 tons of CO2, 2.2 tons of NOx, and 2.6 tons of SO2 per MW reduction. 
  \item  Favoring energy policies related to renewable and low-carbon energy sources from Brazil, Argentina, and Chile are studied.
\end{itemize}

\nomenclature{IEC}{International Electrotechnical Commission}
\nomenclature{FSPV}{Floating solar photovoltaics}
\nomenclature{LBPV}{Land based photovoltaics}
\nomenclature{PV}{Photovoltaics}
\nomenclature{GHI}{Global horizontal irradiance}
\nomenclature{LAC}{Latin America and the Caribbeans}
\nomenclature{LCOE}{Levelized cost of energy}
\nomenclature{NPV}{Net present value}
\nomenclature{IRR}{Internal rate of return}
\nomenclature{PBP}{Payback period}
\nomenclature{O\&M}{Operation and maintenance}
\nomenclature{CAPEX}{Capital Expenditure}

\printnomenclature
\newpage
\section{Introduction}

Access to reliable, affordable, and sustainable electricity remains a critical challenge for many developing economies, where electricity availability is closely linked to economic growth, healthcare, education, industrial development, and overall quality of life. Despite significant progress in global electrification, approximately 750 million people lacked access to electricity in 2023, while nearly 645 million people may still remain without access by 2030 if current trends continue \cite{ref1}. These projections highlight the ongoing challenge of achieving Sustainable Development Goal 7 (SDG7), which aims to ensure universal access to affordable, reliable, sustainable, and modern energy services \cite{ref2}. Although Latin America and the Caribbean (LAC) have achieved an overall electrification rate exceeding 97\%, important disparities remain in electricity reliability, affordability, and service quality, particularly in rural and geographically isolated communities \cite{ref3}. Consequently, future energy strategies in the region must focus not only on expanding generation capacity but also on improving energy security, system resilience, and long-term sustainability \cite{ref4, ref5}.
   
Solar photovoltaic (PV) technology has emerged as a cornerstone of the global energy transition. However, large-scale deployment of conventional ground-mounted PV systems is frequently constrained by land availability, competing land-use priorities, and environmental considerations. Utility-scale PV installations typically require approximately 15,000 m² of land per megawatt of capacity, creating conflicts with agriculture, urban development, and ecological conservation \cite{ref6}. Additional challenges such as complex terrain, fragmented land ownership, elevated operating temperatures, and dust accumulation can further affect project feasibility and performance \cite{ref10, ref7}. Floating Solar Photovoltaic (FSPV) systems have emerged as a promising alternative that can overcome several of these limitations. By using lakes, reservoirs, industrial ponds, and canals as deployment platforms, FSPV systems can expand renewable energy generation while minimizing land-use conflicts.   

The Latin America and the Caribbean (LAC) region represents a particularly attractive frontier for FSPV deployment due to its abundant freshwater resources, extensive hydropower infrastructure, and favorable solar resource availability. When normalized by available water-surface area, the region possesses one of the highest floating solar generation potentials globally. Despite these advantages, FSPV deployment across Latin America and the Caribbean remains limited compared with the rapid commercial expansion observed in several Asian countries. This disparity highlights the need for comprehensive assessments that evaluate the technical, environmental, and deployment potential of FSPV systems within the regional context.
The following subsections, \S\ref{FSPV_concept} through \S\ref{FSPV_objectives}, discuss the concept and advantages of FSPV systems, the motivation for utilizing water bodies in energy-constrained regions, the research gaps addressed by this study, and the principal objectives and contributions of the work.

    \subsection{Concept and advantages of Floating Solar Photovoltaic (FSPV) systems}
    \label{FSPV_concept}
    Floating Solar Photovoltaic (FSPV) systems have emerged as an innovative alternative to conventional land-based PV installations. These systems consist of PV modules mounted on floating platforms deployed over water bodies such as reservoirs, lakes, and canals, with anchoring systems ensuring stability and electrical connections routed to onshore infrastructure \cite{ref11}. 
    FSPV systems offer several advantages. First, they eliminate the need for large land areas, thereby avoiding land-use conflicts and enabling deployment in densely populated or resource-constrained regions \cite{ref8}. Second, the cooling effect of water reduces module temperatures, improving system efficiency and energy yield compared to ground-mounted PV systems \cite{ref9, ref12, ref13, ref14}.
    In addition, FSPV systems contribute to water resource management by reducing evaporation losses through surface shading often by up to 60\% depending on coverage \cite{ref15, ref16, ref17}.
    This benefit is particularly significant in water-stressed regions, where water conservation is critical \cite{ref18, ref19}. Environmental co-benefits include reduced algal growth and improved water quality \cite{ref20}. Furthermore, FSPV systems experience lower dust accumulation, enhancing reliability and reducing maintenance requirements.
    A key advantage of FSPV lies in its compatibility with existing hydropower infrastructure. Hybrid hydro-FSPV systems enable shared transmission infrastructure, improved grid flexibility, and optimized resource utilization, making them particularly attractive for regions with established hydropower capacity \cite{ref21}.
    
    \subsection{Motivation for using water bodies in energy-poor areas}
    \label{FSPV_motivation}
The growing interest in FSPV systems is closely linked to the increasing pressure on energy, land, and water resources in developing regions. While many countries have improved electrification rates over the past two decades, challenges related to electricity reliability, affordability, and long-term energy security continue to persist. In such regions, large lakes, reservoirs, and other inland water bodies represent underutilized assets that can support renewable-energy deployment without creating additional competition for land resources \cite{ref22}. \\
Modern energy-access frameworks increasingly recognize that sustainable energy development requires reliable and resilient power systems capable of supporting economic growth, industrial development, healthcare, education, and other essential services. Consequently, future energy solutions must address not only generation requirements but also broader challenges associated with infrastructure utilization, resource management, and long-term system sustainability.
This opportunity is particularly relevant in Latin America and the Caribbean, where abundant freshwater resources coexist with growing electricity demand and substantial hydropower dependence. Hydropower remains a major contributor to electricity generation across several countries in the region; however, prolonged droughts and changing precipitation patterns have demonstrated the vulnerability of hydropower-dependent systems to climate variability \cite{ref25, ref26}. These challenges have increased interest in complementary renewable-energy technologies capable of utilizing existing infrastructure while improving generation diversity and system resilience. FSPV systems offer a particularly attractive solution because they can be deployed directly on hydropower reservoirs and integrated with existing transmission networks. Hybrid hydro–FSPV systems enable the sharing of substations, transmission infrastructure, and operational facilities while enhancing renewable-energy utilization \cite{ref27}. During periods of high solar generation, hydropower resources can be conserved, thereby improving reservoir management and reducing exposure to hydrological variability \cite{ref28, ref29}. \\
The motivation for investigating FSPV deployment is especially strong in Nicaragua, Honduras, and Guyana. Nicaragua possesses extensive freshwater resources and favorable solar conditions that support renewable-energy expansion. Honduras relies significantly on hydropower generation and therefore presents opportunities for hybrid hydro–FSPV development. Guyana, meanwhile, continues to pursue energy diversification strategies aimed at reducing dependence on fossil-fuel-based electricity generation. Collectively, these countries provide representative examples of how floating solar systems can support energy security, renewable-energy integration, and sustainable resource utilization under different energy-system conditions.

\subsection{Research gap and motivation for South America}\label{FSPV_objectives}



Floating Solar Photovoltaic (FSPV) systems have attracted increasing research interest due to their potential to expand renewable-energy generation while reducing land-use conflicts. Existing studies have investigated various aspects of FSPV deployment, including technical performance, economic feasibility, environmental impacts, and water-conservation benefits. Significant progress has also been made in understanding hybrid hydro–FSPV systems and large-scale deployment opportunities, particularly in Asia where commercial implementation has advanced rapidly.
    
Despite these developments, several important research gaps remain. First, the majority of published FSPV studies are concentrated in Asia, while comparatively limited attention has been devoted to Latin America and the Caribbean despite the region possessing extensive freshwater resources, favorable solar conditions, and substantial hydropower infrastructure.

Second, existing investigations often examine technical performance, environmental impacts, economic feasibility, or policy considerations independently. Relatively few studies integrate these dimensions within a unified framework capable of assessing the broader role of FSPV systems in supporting energy-system development.

Third, comparative assessments between floating solar and conventional land-based photovoltaic systems under similar climatic conditions remain limited. Such comparisons are essential for quantifying the actual performance advantages of floating installations and determining their suitability for large-scale deployment.

These research gaps are particularly relevant for Nicaragua, Honduras, and Guyana. Although these countries possess favorable conditions for floating solar deployment, comprehensive assessments integrating waterbody suitability, technical performance, comparative evaluation with land-based PV systems, hydropower integration opportunities, and deployment feasibility remain scarce.
To address these limitations, the present study develops an integrated techno-socio-economic framework for evaluating FSPV deployment across selected water bodies in Nicaragua, Honduras, and Guyana. The study combines resource assessment, technical performance evaluation, comparative analysis with land-based PV systems, hydropower integration opportunities, and deployment considerations within a unified framework.

\subsection{Objectives and contributions of the paper}



    \label{FSPV_contributions}
    The specific objectives of this study are:
    \begin{enumerate}
    \item To assess the technical potential of FSPV deployment across major water bodies in Nicaragua, Honduras, and Guyana using realistic deployment footprints and utility-scale development scenarios.
    \item To evaluate the long-term performance of representative FSPV systems through detailed analysis of annual energy generation, energy yield, performance ratio, and capacity factor.
    \item To compare the performance of floating solar systems with equivalent land-based photovoltaic (LBPV) installations under identical regional climatic conditions.
    \item To investigate integration pathways for hybrid hydro–FSPV systems and assess the associated benefits regarding transmission sharing, grid connectivity, and renewable-energy utilization.
    \item To examine the potential role of FSPV systems in supporting energy security, renewable-energy expansion, and sustainable resource utilization within the Latin America and Caribbean (LAC) region.
\end{enumerate}

The principal contributions of this work are summarized as follows:
\begin{enumerate}
    \item Development of an integrated techno-socio-economic framework that combines resource assessment, technical performance analysis, hydropower integration opportunities, and deployment considerations for FSPV systems in the LAC region.
    \item Quantification of the floating solar potential of strategically important water bodies in Nicaragua, Honduras, and Guyana through detailed location-specific case studies.
    \item Comprehensive comparative assessment of FSPV and LBPV systems under representative climatic conditions, providing insights into the operational advantages of floating solar deployment.
    \item Evaluation of hybrid hydro–FSPV deployment opportunities and their potential role in improving renewable-energy utilization, infrastructure sharing, and energy-system resilience.
    \item Providing policy-relevant recommendations to support future planning, investment, and deployment of floating solar systems across emerging economies in Latin America and the Caribbean.
    \item Establishment of a replicable assessment framework that can be applied to other regions possessing significant freshwater resources and untapped floating solar potential.
\end{enumerate}

\section{Literature Review and Research Gap Analysis}
The growing demand for reliable, affordable, and sustainable electricity has accelerated research on renewable-energy technologies and modern electrification strategies. Among these technologies, Floating Solar Photovoltaic (FSPV) systems have emerged as a promising solution capable of addressing land constraints, improving renewable-energy utilization, and supporting energy-system resilience. Existing studies have examined FSPV deployment from technical, environmental, economic, and policy perspectives. However, the literature remains fragmented, particularly regarding the integration of energy-access challenges, hydropower dependency, water-resource utilization, and deployment feasibility within emerging economies. The following subsections review current developments in energy-access frameworks, regional electrification challenges, and floating solar technologies, leading to the identification of research gaps relevant to South America.

\subsection{Marginalized Energy Communities and Electricity Access Frameworks}\label{Marginalized_Energy_Communities}
Energy poverty is increasingly recognized as a multidimensional challenge that extends beyond the simple availability of electricity. Contemporary frameworks define energy poverty through a combination of accessibility, affordability, reliability, and quality of energy services, reflecting the broader role of energy in supporting human development and socioeconomic well-being \cite{ref39, ref40}. Traditional electrification metrics frequently classify households as electrified based solely on grid connectivity, even when electricity supply remains unreliable, unaffordable, or insufficient to support productive activities. To address these limitations, multidimensional assessment approaches such as the Multidimensional Energy Poverty Index (MEPI) were developed to capture a broader range of household energy deprivations \cite{ref41}. Subsequent studies have further emphasized the importance of integrated frameworks for evaluating energy poverty across diverse geographical and socioeconomic contexts \cite{ref42}. \\
The consequences of energy poverty extend far beyond the energy sector itself. Limited access to reliable electricity constrains educational attainment, healthcare delivery, communication, and economic productivity, thereby reinforcing existing development challenges \cite{ref43}. Empirical studies have demonstrated significant associations between energy deprivation and adverse health outcomes, including respiratory illnesses, increased mortality risks, and reduced quality of life \cite{ref44}. These impacts are particularly severe in communities that remain dependent on traditional biomass fuels for cooking and heating, where prolonged exposure to indoor air pollution creates substantial public-health concerns \cite{ref45}.\\
Energy poverty also exhibits a strong social dimension. Women and vulnerable populations are often disproportionately affected by inadequate energy access due to their greater exposure to household energy-related activities and limited access to educational and economic opportunities \cite{ref46}. Consequently, energy deprivation contributes to the persistence of broader social inequalities, highlighting the need for inclusive and equitable energy policies \cite{ref48, ref47}. These findings collectively demonstrate that energy poverty is not solely a technical challenge but also a social and developmental issue requiring integrated policy interventions. \\
Although substantial progress has been achieved in global electrification, evidence increasingly suggests that access alone is insufficient for achieving sustainable development objectives. Reliable, affordable, and resilient electricity systems are essential for supporting long-term economic growth and improving living standards. Consequently, growing attention has shifted toward renewable-energy technologies capable of delivering sustainable electricity access while minimizing environmental impacts. Within this context, renewable-energy systems, including emerging technologies such as Floating Solar Photovoltaic (FSPV), offer opportunities to support energy-access objectives while simultaneously addressing resource constraints and broader sustainability goals.

\subsection{Regional Perspectives on Latin American Energy Access and Grid Vulnerabilities}\label{Regional_Perspectives}
Although Latin America and the Caribbean (LAC) have achieved relatively high national electrification rates compared with many developing regions, significant disparities in electricity access, reliability, and service quality persist across countries and communities. National electrification statistics often conceal substantial sub-national inequalities, particularly in rural, geographically isolated, and low-income regions where electricity services remain unreliable or insufficient to support socioeconomic development (Mori, 2016). Consequently, achieving universal electrification requires not only expanding access but also improving the quality, affordability, and resilience of electricity supply systems. \\
The region faces additional challenges associated with aging infrastructure, increasing electricity demand, and growing exposure to climate-related risks. Several countries remain highly dependent on hydropower resources, making electricity systems vulnerable to prolonged droughts, changing precipitation patterns, and hydrological variability. These vulnerabilities have increased the need for diversified electricity-generation portfolios capable of improving system resilience while supporting long-term sustainability objectives. \\
In response to these challenges, decentralized renewable-energy technologies have emerged as important tools for improving electricity access in underserved communities. Solar home systems, mini-grids, and pico-photovoltaic technologies have demonstrated considerable success in extending electricity access to remote regions while generating broader socioeconomic benefits, including increased household income, improved educational outcomes, and reduced dependence on conventional fuels \cite{ref55, ref54}. These experiences illustrate the potential of renewable-energy technologies to address both energy-access and development challenges simultaneously. \\
However, the long-term effectiveness of electrification initiatives depends on more than technology deployment alone. Previous studies have emphasized the importance of supportive policy frameworks, sustainable financing mechanisms, institutional capacity, and coordinated planning approaches for ensuring successful implementation and long-term operational viability \cite{ref57}. Integrated strategies that combine centralized grid expansion with decentralized renewable-energy solutions have consistently been identified as the most cost-effective pathways toward universal electrification \cite{ref58}.\\
While decentralized systems play a critical role in improving community-level access, growing electricity demand and regional decarbonization goals also require large-scale renewable-energy deployment. In this context, technologies capable of utilizing existing infrastructure while minimizing land-use conflicts are becoming increasingly important. For Latin America and the Caribbean, where abundant freshwater resources coexist with extensive hydropower infrastructure, Floating Solar Photovoltaic (FSPV) systems represent a promising opportunity for expanding renewable-energy generation, enhancing system resilience, and supporting broader energy-transition objectives.
Consequently, understanding the deployment potential and operational implications of FSPV systems within the regional context has become increasingly important for future energy planning. This is particularly relevant for emerging economies seeking to balance renewable-energy expansion, energy security, and sustainable resource utilization under changing climatic and economic conditions.

\subsection{Comprehensive Performance Transitions: Land-Based Solar, FSPV, and Hybrid Configurations}\label{Comprehensive_Performance_Transitions}
Solar photovoltaic (PV) technology has become one of the most rapidly expanding renewable-energy technologies worldwide. Despite significant technological advancements and declining installation costs, conventional land-based photovoltaic (LBPV) systems continue to face several technical, environmental, and socioeconomic challenges. One of the most significant limitations is land availability, as utility-scale PV installations require extensive land areas that often compete with agricultural production, urban expansion, and environmental conservation objectives \cite{ref6, ref8}. These challenges are particularly pronounced in regions characterized by fragmented land ownership, limited suitable terrain, and increasing pressure on land resources \cite{ref35}.\\
Operational performance also represents an important concern for LBPV systems. Elevated module operating temperatures reduce photovoltaic conversion efficiency and increase thermal losses, particularly in tropical and subtropical environments\cite{ref9, ref14}. Dust accumulation and soiling further decrease energy generation and increase maintenance requirements, negatively affecting long-term system performance and project economics \cite{ref7}. Collectively, these limitations have encouraged the exploration of alternative deployment approaches capable of improving resource utilization while maintaining high renewable-energy output.\\
Floating Solar Photovoltaic (FSPV) systems have emerged as a promising solution to many of these challenges. By utilizing lakes, reservoirs, ponds, and other water bodies as deployment platforms, FSPV systems eliminate direct competition for land resources while enabling large-scale renewable-energy development \cite{ref32}. Several studies have demonstrated substantial technical potential for floating solar deployment in countries such as Brazil and Chile, where even partial reservoir coverage can generate significant amounts of electricity \cite{ref30, ref31}. The cooling effect provided by the underlying water body has also been shown to reduce module operating temperatures and improve energy-generation performance relative to equivalent land-based systems \cite{ref9, ref13, ref14, ref12}.\\
Beyond improvements in electrical performance, FSPV systems provide additional environmental and operational benefits. Previous investigations have reported reductions in water evaporation, improved water-resource management, and enhanced utilization of existing infrastructure \cite{abdelhady2021performance, ref16, ref15, ref18, ref19, ref20}. Economic assessments have further suggested that FSPV systems can achieve competitive levelized costs of electricity and attractive financial returns under appropriate deployment conditions \cite{ref59}. These advantages have contributed to growing interest in floating solar technologies as a means of supporting renewable-energy expansion while minimizing land-use conflicts.
Nevertheless, FSPV deployment is associated with several technical and environmental challenges. Floating structures are exposed to varying wind, wave, and water-level conditions that influence anchoring requirements, structural stability, and long-term operational reliability \cite{ref38, ref13, ref14}. Potential ecological impacts and maintenance considerations also require careful assessment to ensure sustainable implementation \cite{ref37, ref35}. Consequently, successful deployment depends on site-specific engineering design, environmental evaluation, and long-term operational planning.\\
Recent research has increasingly focused on hybrid hydro–FSPV configurations as a pathway toward enhanced renewable-energy integration. Hydropower reservoirs provide attractive deployment locations because they offer existing transmission infrastructure, substations, and grid interconnections that can be shared between technologies \cite{ref24, ref33, ref23, ref21}. Hybrid systems can improve infrastructure utilization, enhance operational flexibility, and increase renewable-energy penetration while reducing exposure to hydrological variability. During periods of high solar generation, hydropower resources can be conserved, thereby improving reservoir management and strengthening overall system resilience \cite{ref30, ref28}.
A comparative summary of representative studies on LBPV systems, FSPV technologies, and hybrid hydro–FSPV configurations is presented in Table \ref{tab:fspv_comparison}. The reviewed literature consistently demonstrates that FSPV systems offer significant advantages in terms of land-use efficiency, thermal performance, water conservation, and infrastructure sharing. At the same time, the studies reveal persistent uncertainties related to long-term environmental impacts, site-specific deployment requirements, economic feasibility under varying regional conditions, and comparative performance relative to conventional land-based PV systems.
Overall, the transition from conventional land-based PV installations toward floating and hybrid solar configurations reflects a broader shift toward integrated energy and resource management approaches. While existing studies have substantially advanced understanding of FSPV technologies, important questions remain regarding deployment strategies, regional suitability, comparative system performance, and large-scale implementation pathways. These unresolved issues are particularly relevant for Latin America and the Caribbean, where abundant freshwater resources and extensive hydropower infrastructure create favorable conditions for future floating solar development.

\begin{table}[t]
\small
\centering
\caption{Comparative analysis of energy access solutions and FSPV systems}
\label{tab:fspv_comparison}
\begin{tabular}{p{2.5cm} p{2.5cm} p{3cm} p{3cm} p{3.5cm}}
\hline
\textbf{Category} &
\textbf{Key aspect} &
\textbf{Land based PV} &
\textbf{Floating solar PV}  &
\textbf{Hybrid Hydro-FSPV}\\
\hline

Energy access role &
Suitability for rural/energy-poor regions &
Effective but limited by land availability and infrastructure constraints &
Highly suitable due to deployment on existing water bodies &
Enhances reliability by complementing hydropower variability\\

Energy access role &
Off-grid applicability &
Used in decentralized systems (mini-grids, SHS) \cite{ref54, ref55} &
Emerging application in reservoir-based electrification \cite{ref22} &
Supports grid-connected systems with improved flexibility \cite{ref27, ref28}\\

Land requirement &
Land usage &
Requires 15,000 m²/MW, leading to land-use conflicts \cite{ref6, ref7, ref8, ref9} &
No land requirement; utilizes water surfaces \cite{ref6, ref7} &
No additional land requirement due to co-location \cite{ref22}\\

Geographical constraints &
Terrain limitations &
Challenging in hilly, forested, or fragmented land regions \cite{ref10} &
Less affected by terrain constraints &
Uses existing hydropower sites, reducing siting complexity \cite{ref28}\\

Energy efficiency &
Thermal performance &
Reduced efficiency due to high temperatures \cite{ref6, ref7} &
Improved efficiency due to water cooling (11\% gain) \cite{ref12, ref9} &
Enhanced system-level efficiency through hybrid operation \cite{ref21, ref22}\\

Water resource impact &
Evaporation effects &
No impact on water conservation &
Reduces evaporation up to 60\% \cite{ref15, ref16, ref17} &
Supports water-energy nexus; improves reservoir management \cite{ref18, ref19, ref26}\\

Environmental impact &
Land/ecosystem effects &
Potential biodiversity loss and land-use conflicts \cite{ref10} &
May affect aquatic ecosystems (reduced light penetration, oxygen levels) \cite{ref5, ref6, ref7} &
Lower impact than new dams or thermal plants \cite{ref29}\\

Economic performance &
Cost and feasibility &
Competitive but influenced by land costs &
LCOE \$32–55/MWh; positive NPV in many cases \cite{ref59} &
Reduced payback due to shared infrastructure \cite{ref33} [60]\\

Infrastructure needs &
Grid integration &
Requires new transmission infrastructure in remote areas &
Can be located near demand centers \cite{ref15} &
Uses existing hydropower infrastructure, reducing costs \cite{ref28}\\

\hline
\end{tabular} 
\end{table}

\subsection{Literature Synthesis and Remaining Research Challenges}\label{synthesis}
The literature reviewed in sections \S\ref{Marginalized_Energy_Communities} through \S\ref{Comprehensive_Performance_Transitions} demonstrates the growing importance of renewable-energy technologies in addressing electricity-access challenges while supporting broader sustainability objectives. Studies on energy poverty consistently emphasize that reliable, affordable, and resilient electricity systems are essential for improving socioeconomic development, healthcare, education, and overall quality of life \cite{ref39, ref40, ref43}. At the same time, investigations into decentralized electrification pathways highlight the significant role of renewable-energy technologies in extending electricity access to underserved and geographically isolated communities \cite{ref55, ref54, ref58}.\\
The reviewed literature further demonstrates that Floating Solar Photovoltaic (FSPV) systems provide several advantages over conventional land-based photovoltaic installations, including improved land-use efficiency, enhanced thermal performance, water-conservation benefits, and opportunities for infrastructure sharing \cite{ref6, ref9, abdelhady2021performance}. Moreover, hybrid hydro–FSPV configurations have emerged as promising solutions for increasing renewable-energy utilization, improving operational flexibility, and enhancing energy-system resilience through the integration of complementary renewable resources \cite{ref21, ref23, ref28}.\\
As summarized in Table \ref{tab:fspv_comparison}, existing studies have substantially advanced the understanding of FSPV technologies from technical, environmental, economic, and policy perspectives. However, the literature remains fragmented in several important respects. Most investigations focus on individual dimensions of FSPV deployment rather than adopting integrated assessment approaches capable of simultaneously evaluating resource availability, technical performance, comparative system behavior, hydropower integration opportunities, and deployment feasibility. Consequently, the broader role of FSPV systems within evolving energy systems remains insufficiently understood.
Another notable observation is the limited availability of comprehensive studies focused on Latin America and the Caribbean. Despite the region possessing abundant freshwater resources, extensive hydropower infrastructure, and favorable solar-energy potential, the majority of FSPV research remains concentrated in Asia and selected developed economies. As a result, uncertainties persist regarding the practical deployment potential, long-term performance, and broader energy-system implications of FSPV systems within many countries of the region.
These remaining challenges highlight the need for integrated evaluation frameworks capable of combining technical, environmental, and energy-system perspectives within a unified assessment approach. Such frameworks are particularly relevant for countries such as Nicaragua, Honduras, and Guyana, where renewable-energy expansion, hydropower utilization, and energy-access objectives remain closely interconnected. Addressing these challenges requires not only the assessment of floating solar potential but also an improved understanding of comparative system performance, hydropower integration opportunities, and deployment pathways under region-specific conditions.
Motivated by these considerations, the present study develops a comprehensive techno-socio-economic framework for evaluating FSPV deployment opportunities across selected water bodies in Nicaragua, Honduras, and Guyana. By integrating resource assessment, technical performance evaluation, comparative analysis with land-based PV systems, and hydropower integration considerations within a single framework, the study seeks to contribute to the growing body of knowledge on floating solar deployment in Latin America and the Caribbean.

\section{Global developments in floating solar photovoltaics (FSPV) and the case for South America}

\subsection{Global scenario}
\begin{figure*}[!htbp]
    \centering
    \includegraphics[width=0.6\textwidth]{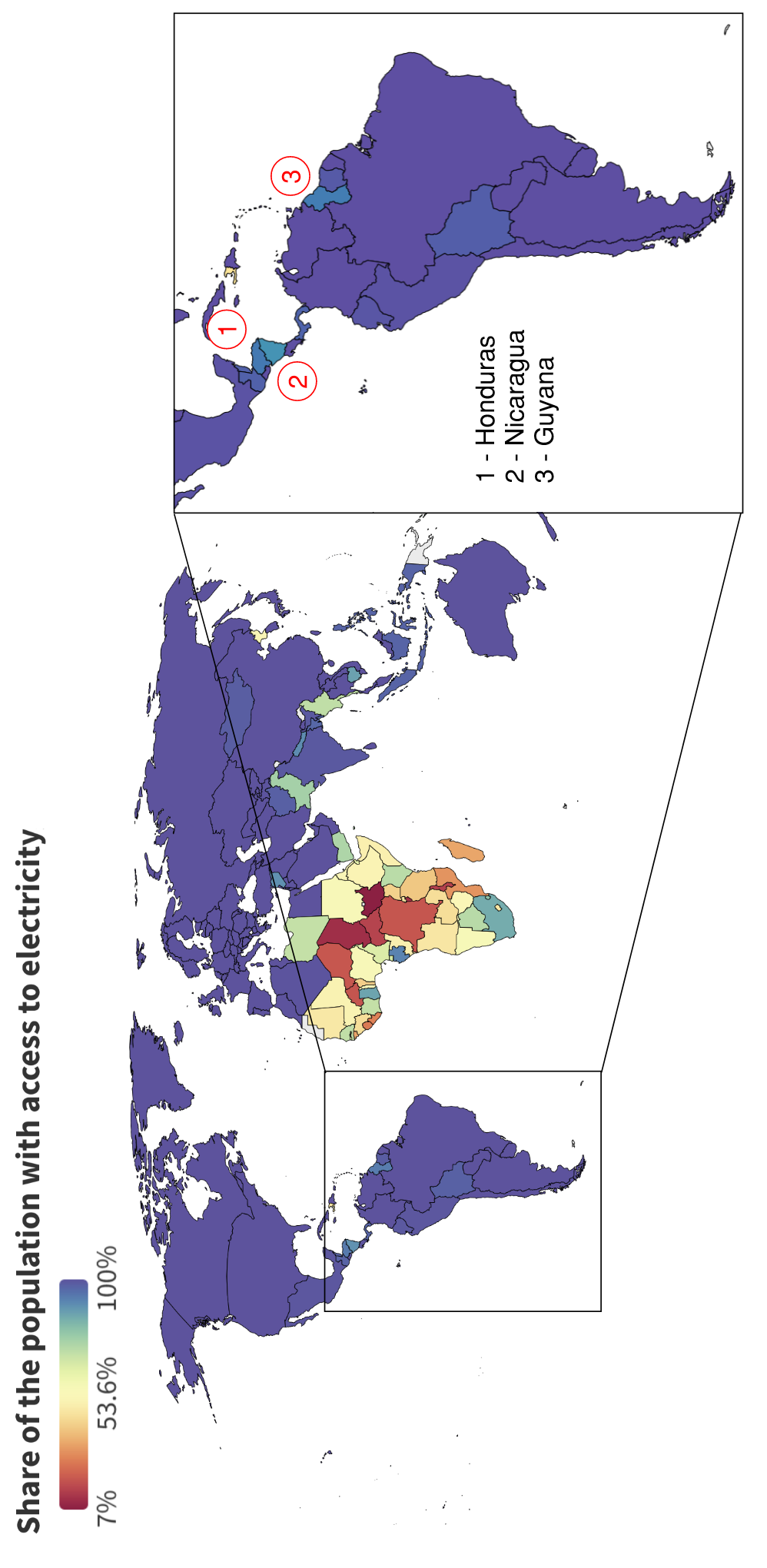}
    \caption{Share of the population with access to electricity. Continents with 90\% or less share of the population with access to electricity are Africa, Asia, and South America. }
    \label{fig:threecountries}
\end{figure*}

The growing global demand for reliable, affordable, and sustainable electricity has accelerated the deployment of renewable-energy technologies across both developed and developing economies. Despite significant improvements in electrification rates over recent decades, substantial disparities in electricity access continue to persist worldwide. Figure \ref{fig:threecountries} illustrates the global distribution of electricity access and highlights marked regional inequalities in energy availability. While most countries in North America, Europe, and large parts of Latin America have achieved relatively high electrification rates, several regions in Sub-Saharan Africa and parts of Asia continue to experience significant access deficits. The figure demonstrates that many countries still contain marginalized communities where reliable electricity access remains below desirable levels, emphasizing the continuing need for scalable and sustainable energy solutions capable of supporting socioeconomic development and long-term energy security.\\

Among the various renewable-energy technologies available, Floating Solar Photovoltaic (FSPV) systems have emerged as a particularly attractive option due to their ability to generate electricity without competing directly for valuable land resources. By utilizing reservoirs, lakes, ponds, and other water bodies, FSPV systems can support renewable-energy expansion while simultaneously providing benefits such as reduced water evaporation, improved thermal performance, and enhanced utilization of existing infrastructure \cite{ref18, ref19}. These advantages have stimulated growing interest in floating solar technologies across regions characterized by increasing electricity demand and limited land availability.\\

The global technical potential of FSPV deployment has been evaluated across multiple continents and is summarized in Table \ref{tab:fspv_potential_three}. The results reveal substantial differences in both absolute and normalized generation potential. Asia possesses the largest theoretical annual generation potential, estimated at approximately 3,754 TWh, reflecting its extensive freshwater resources and large number of suitable reservoirs. North America and Africa also demonstrate considerable generation potential, with estimated capacities of approximately 1,842 TWh and 1,743 TWh, respectively. However, a more meaningful comparison emerges when generation potential is normalized by available water-surface area. As shown in Table 2, Asia exhibits a normalized generation potential of approximately 18.15 TWh per million acres, while Africa achieves approximately 11.59 TWh per million acres. In contrast, South America demonstrates the highest normalized generation potential globally, reaching approximately 38.26 TWh per million acres. This value is more than double that of Asia and more than three times that of Africa, indicating that South American water bodies offer exceptionally favorable conditions for floating solar deployment. These findings provide a strong regional justification for investigating FSPV opportunities within Central and South America. \\

\begin{table}[t]
\small
\centering
\caption{FSPV potential distribution in lakes and reservoirs across Asia, Africa, and South America.*}
\label{tab:fspv_potential_three}
\renewcommand{\arraystretch}{1.2} 
\begin{tabular}{p{2.25cm} p{3cm} p{2cm} p{4cm} p{6cm}}
\hline
\textbf{Geographic region} &
\textbf{Total surface area (million acres)} &
\textbf{Number of waterbodies evaluated} &
\textbf{Yearly energy generation potential (TWh) when 10\% of the suitable water surface area is covered with FSPV}  \\
\hline

Asia &
206.74 &
155,878 &
3,754\\

Africa &
53.82 &
13,584 &
624\\

South America &
32.82 &
53,825 &
1,256\\

\hline
\end{tabular}
\\[3pt]
\footnotesize\textsuperscript{*}Source: authors’ compilation derieved from  \cite{woolway2024decarbonization}. See Data Availability Statement at the end of this manuscript for link to data. 
\end{table}

While resource availability establishes technical potential, the commercial maturity of FSPV technology is reflected through existing deployment experience. Table \ref{tab:fspv_global_summary} summarizes representative operational and planned FSPV projects across Asia, Africa, and South America. The table illustrates the rapid expansion of floating solar systems from small pilot installations to utility-scale projects exceeding several hundred megawatts. Asia currently dominates global deployment, hosting some of the world's largest FSPV installations, including the 320 MW Dezhou Dingzhuang Floating Solar Farm in China as well as large-scale developments in India, Indonesia, Thailand, Singapore, and South Korea. These projects demonstrate the technical feasibility, scalability, and commercial viability of floating solar technologies under diverse climatic and operational conditions (Silva et al., 2023a; Silva et al., 2023b).\\
The deployment patterns presented in Table 3 also reveal notable differences in market maturity across regions. In Africa, FSPV development remains in its early stages, with projects such as the Bui Floating Solar Farm in Ghana representing important pilot-scale initiatives. South America presents an interesting contrast. Despite possessing the highest normalized generation potential globally, commercial deployment remains comparatively limited, with notable examples including the Aquasol project in Colombia and the floating solar installation on the Lajeado Reservoir in Brazil. This disparity between resource availability and implementation highlights a significant opportunity for future expansion and underscores the importance of region-specific studies capable of translating continental-scale potential into practical deployment strategies. \\
Collectively, Figure \ref{fig:threecountries}, Table \ref{tab:fspv_potential_three}, and Table \ref{tab:fspv_global_summary} provide complementary evidence supporting the strategic relevance of FSPV technology. Figure \ref{fig:threecountries} highlights the continuing need for sustainable electricity solutions in underserved regions, Table \ref{tab:fspv_potential_three} demonstrates the exceptional floating solar potential of South America, and Table \ref{tab:fspv_global_summary} confirms the technical maturity and global scalability of the technology, by summarizing major FSPV projects across Africa, Asia, and South America. Together, these findings establish a strong foundation for focusing on Central and South America, where abundant freshwater resources and favorable solar conditions create significant opportunities for future FSPV development.

\begingroup
\fontsize{9}{11}\selectfont
\begin{longtable}{p{2.25cm} p{2.5cm} p{1.2cm} p{1.5cm} p{5.5cm} p{2cm}}
    
    \caption{Summary of major FSPV projects across Africa, Asia, and South America. \\ 
    }\\
    \label{tab:fspv_global_summary} \\
    \hline
    \textbf{Country} &
    \textbf{Major FSPV project} &
    \textbf{Capacity} &
    \textbf{Water body type} &
    \textbf{Key details} &
    \textbf{References} \\
    \hline
    \endfirsthead

    \hline
    \multicolumn{6}{r}{End of the table.}\\
    \hline
    \endlastfoot

    \hline
    \multicolumn{6}{l}{Continued from the previous page.}\\
    \hline
    \textbf{Country} &
    \textbf{Major FSPV project} &
    \textbf{Capacity} &
    \textbf{Water body type} &
    \textbf{Key details} &
    \textbf{References} \\
    \hline
    \endhead

    \hline
    \multicolumn{6}{r}{Continued on the next page \ldots}\\
    \hline
    \endfoot

    Asia - China &
    Dezhou Dingzhuang floating solar farm &
    320 MW &
    Reservoir/ flooded mining area &
    Located in Shandong Province on a former aquaculture pond, this is one of the world's largest operational floating solar projects. The project uses bifacial solar modules to increase energy yield by capturing reflected light from the water surface and incorporates a real-time monitoring system for mooring line tension and inverter performance. Annual generation is estimated at approximately 400 GWh, offsetting 320,000 tonnes of CO2. &
    \cite{huang2023comprehensive}\\

    Asia - China &
    Huainan floating solar farm, Anhui &
    40 MW &
    Flooded coal mine &
    Built on an abandoned coal mining area in Anhui Province, this project demonstrates how flooded mine pits can be converted into renewable energy assets while reducing land-use conflicts. Water depth ranges from 4 to 10 meters, and the floating platform uses high-density polyethylene (HDPE) floats designed to withstand typhoon winds up to 160 km/h. The project includes water quality monitoring stations to track heavy metal concentrations post-installation. &
    \cite{wang2022influence}\\

    Asia - India &
    NTPC Ramagundam floating solar plant &
    100 MW &
    Reservoir &
    Located in Telangana and covering 450 acres, this is one of India's largest operational FSPV plants. Commissioned in July 2022, it consists of 40 blocks, each of 2.5 MW with independent inverters and floating transformers. The project reduces evaporation losses by an estimated 1.3 billion litres of water annually and includes a fish-friendly design allowing sunlight penetration for aquatic life. &
    \cite{charles2023floating}\\

    Asia - India &
    NTPC Haripad / Kayamkulam floating solar plant &
    92 MW &
    Reservoir &
    Constructed on reservoirs in Kerala, spanning 350 acres. Commissioned in three phases in 2022, the plant uses a combination of polycrystalline and monocrystalline modules. Expected annual generation is 150 GWh, and the project includes a 1 MW battery energy storage system for grid stabilization. &
    \cite{charles2023floating}\\

    Asia - India &
    Omkareshwar floating solar power park &
    600 MW &
    Dam reservoir &
    Located in West Java, this is one of Southeast Asia's largest FSPV projects. Commissioned in November 2023, it uses ~340,000 solar panels across 250 hectares. It is integrated with the 1,008 MW Cirata hydropower plant, allowing shared transmission infrastructure. Annual generation is approximately 245 GWh. &
    \cite{gupta2025sustainable}\\

    Asia - Indonesia &
    Cirata floating solar plant &
    145 MWp &
    Reservoir &
    One of the largest planned floating solar parks globally, developed on the Omkareshwar Dam in Madhya Pradesh. Phase I (278 MW) was commissioned in 2024. Post-storm damage in July 2024, the redesign includes reinforced mooring with 12 anchor points per block. Upon full completion, projected annual generation is 1,140 GWh. &
    \cite{rifansyah2024techno}\\

    Asia - Indonesia &
    Saguling  floating solar project &
    92 MWp &
    Reservoir &
    Located on the Saguling hydropower reservoir in West Java. Construction is expected to begin in 2025 with commissioning targeted for 2027. The project will use high-efficiency bifacial modules and is expected to generate more than 130 GWh annually, reducing approximately 104,000 tons of CO2 emissions per year. &
    \cite{kibtiah2024floating}\\
    
    Asia - Singapore &
    Tengeh reservoir floating solar farm &
    60 MWp &
    Reservoir &
    Covering 45 hectares, this is one of the world's largest inland reservoir-based FSPV plants. Commissioned in July 2021, it is integrated with a 1.5 MW/3 MWh battery storage system and includes Singapore's first large-scale floating transformer station. Annual generation is approximately 77 GWh. &
    \cite{bagri2025econo}\\
    
    Asia - Thailand &
    Sirindhorn dam floating solar farm &
    45 MWp &
    Hydropower reservoir &
    Located in Ubon Ratchathani, this hybrid floating solar-hydropower project was commissioned in October 2021. It features a hybrid control system that automatically switches between solar and hydropower based on grid demand and weather. Annual generation is 45 GWh, and it includes a 3 MW/3 MWh lithium-ion battery. &
    \cite{sapthanakorn2021evaluating}\\
    
    Asia - Vietnam &
    Da Mi floating solar plant &
    47.5 MWp &
    Hydropower reservoir &
    Commissioned in May 2019 as Vietnam's first FSPV project. The plant uses 143,472 solar panels covering 35 hectares. It shares inverter stations and the 220 kV transmission line with the hydropower plant. Annual generation is 69 GWh, increasing the reservoir's total renewable output by 15\%. &
    \cite{nguyen2023technical}\\   

    Asia - Japan &
    Yamakura dam floating solar plant &
    13.7 MW &
    Reservoir &
    Located in Chiba Prefecture, commissioned in 2018. Designed to withstand typhoon wind speeds of 180 km/h and seismic activity, it is Japan's first earthquake-certified floating solar array. Annual generation is 16 GWh, and the project includes a 2.5 MW/2.5 MWh battery storage system. &
    \cite{honaryar2022wind}\\  

    Asia - South Korea &
    Saemangeum floating solar plant &
    2100 MW (planned) &
    Seawater basin / tidal flat &
    Located on the Saemangeum seawall area. Phase 1 (300-500 MW) is pending environmental impact assessment (2026-2028). Technology challenges include saltwater corrosion and wave loading, leading to the development of marine-grade structures and salt-resistant modules. &
    \cite{acharya2019floating}\\     

       Africa - Egypt &
    Benban floating solar farm &
    5 MW &
    Developed as a pilot floating solar & installation linked to the larger Benban Solar Park region. The project explores the use of floating PV on irrigation canals and water storage basins to reduce evaporation losses in arid conditions while improving land-use efficiency. The system is designed with corrosion-resistant floats and elevated mounting structures to withstand high temperatures and dust accumulation. &
    \cite{mohamed2022techno}\\

    Africa - Ghana &
    Bui floating solar farm &
    5 MW (phase 1) &
    Hydropower reservoir &
    Installed on the Bui Reservoir, this project was commissioned as one of the first operational floating solar systems in West Africa. It is integrated with the Bui Hydropower Plant and includes plans for future expansion to more than 60 MW. The project demonstrates the potential of hybrid hydro-floating solar systems in improving generation during dry seasons. &
    \cite{asare2024assessing}\\

    Africa - Egypt &
    Aswan Dam Lake Nasser floating solar proposal &
    300 MW (planned) &
    Reservoir &
    Proposed on Lake Nasser near the Aswan High Dam, this project is intended to integrate floating solar with existing hydropower infrastructure. The large reservoir area offers major potential for hybrid hydro-solar generation and improved grid stability. The project is expected to reduce water evaporation and improve renewable energy generation in southern Egypt. &
    \cite{elshafei2021study}\\

    Africa - Kenya &
    Seven Forks dam floating solar project &
    42.5 MW (planned) &
    Hydropower reservoir &
    Proposed across the Seven Forks cascade reservoirs, this project aims to integrate floating solar with Kenya’s hydropower network. The project is expected to reduce seasonal variability in electricity generation and support national renewable energy expansion goals. &
    \cite{kenyafspv}\\

    Africa - Uganda &
    Kariba floating solar project &
    100 MW (planned) &
    Hydropower reservoir &
    Proposed on Lake Kariba, this project aims to address declining hydropower generation caused by drought and fluctuating water levels. Floating solar is expected to provide additional power generation capacity while using existing grid infrastructure connected to the Kariba hydropower station. &
    \cite{chirwa2023floating}\\

    South America - Brazil &
    Lajeado hydropower reservoir in Tocantins &
    54 MW &
    Hydropower reservoir &
    South America has recently witnessed growing interest in FSPV deployment, particularly in countries with strong hydropower infrastructure and increasing renewable energy targets.  &
    \cite{coelho2017comparison}\\

    South America - Columbia &
    YurbaQua floating solar project - Turbaco, Bolívar  &
    2.8 MWp &
    Artificial water reservoir within an industrial free trade zone &
    YurbaQua is sited on an artificial water reservoir within an industrial free trade zone rather than a hydropower reservoir, illustrating that diverse water body types can host FSPV systems across the region. &
     - \\

    South America - Columbia &
    Aquasol project &
    1.5 MW &
    Hydropower reservoir &
    The Aquasol project, a 1.5 MW pilot installation at the 340 MW Urrá hydropower plant on the Sinú River in Córdoba, was commissioned in 2023 and was recognized as the largest floating solar installation at a hydropower reservoir in South America at the time of its launch.   &
    \cite{vargas2023conceptual}\\
    
    \hline
\end{longtable}
\endgroup

\subsection{The case study for South America}

South America represents one of the most promising regions for Floating Solar Photovoltaic (FSPV) deployment due to its extensive freshwater resources, favorable solar conditions, and large hydropower infrastructure. As shown in Table \ref{tab:fspv_potential_three}, the region exhibits the highest normalized floating solar generation potential globally, reaching approximately 38.26 TWh per million acres of water surface area, significantly exceeding Asia (18.15 TWh per million acres) and Africa (11.59 TWh per million acres).
The region also possesses numerous lakes and reservoirs suitable for floating solar deployment. In addition, many South American countries rely heavily on hydropower generation, creating opportunities for hybrid hydro–FSPV systems that can improve infrastructure utilization and enhance system resilience under changing climatic conditions \cite{ref19}.
Despite these advantages, Table \ref{tab:fspv_global_summary} indicates that commercial FSPV deployment in South America remains limited compared with Asia, where several large-scale projects are already operational. This gap between resource availability and implementation highlights the need for region-specific assessments capable of identifying practical deployment opportunities. Therefore, South America provides an appropriate and strategically important region for evaluating the technical potential and energy-system benefits of FSPV technologies.

\subsection{The FSPV argument for Nicaragua, Honduras, and Guyana}
\begin{figure*}[!htbp]
    \centering
    \includegraphics[width=0.9\textwidth]{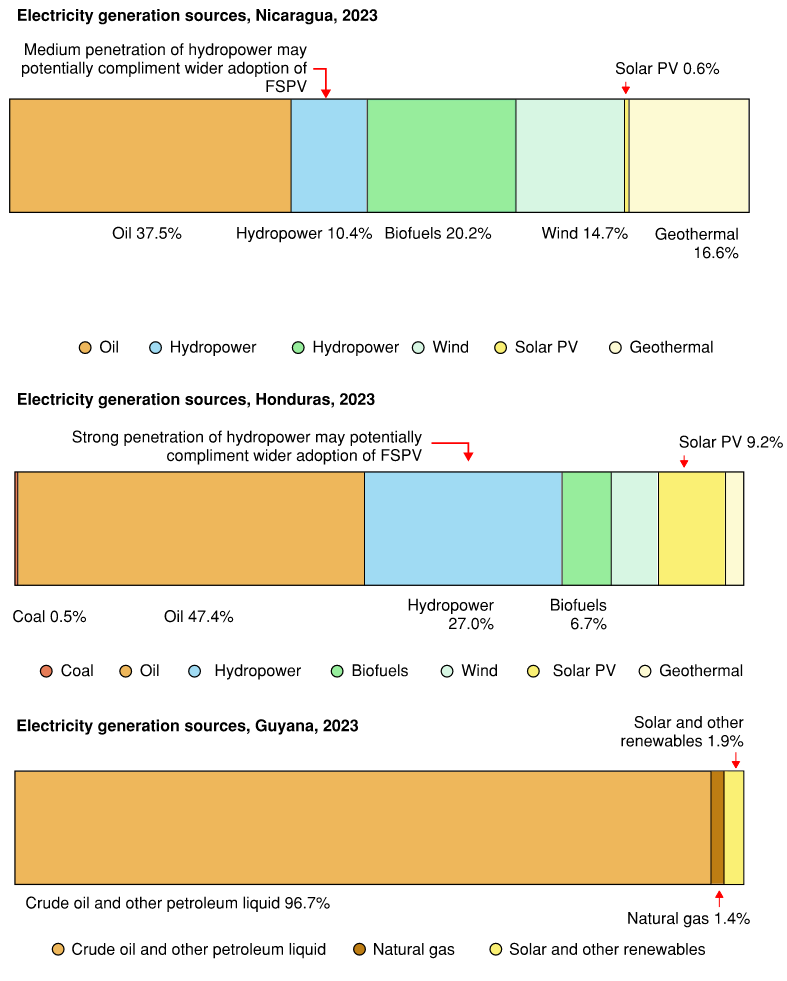}
    \caption{Comparison of electricity generation sources in Nicaragua, Honduras, and Guyana for 2023.}
    
    \label{fig:elegensource}
\end{figure*}

 Nicaragua, Honduras, and Guyana were selected as representative case-study countries due to their contrasting electricity-generation structures, renewable-energy development pathways, and opportunities for floating solar deployment. Figure \ref{fig:elegensource} illustrates the electricity-generation mix of the selected countries and highlights their distinct energy-system characteristics.
Nicaragua has achieved significant renewable-energy penetration through hydropower, wind, geothermal, and bioenergy resources. However, solar energy currently contributes only a small fraction of the national electricity mix, indicating considerable potential for future expansion through FSPV deployment. Honduras presents a different scenario, with hydropower accounting for a significant share of electricity generation. This creates favorable conditions for hybrid hydro–FSPV systems that can utilize existing reservoir infrastructure and transmission networks while improving operational flexibility. In contrast, Guyana remains heavily dependent on fossil-fuel-based electricity generation, highlighting the potential role of FSPV systems in supporting renewable-energy diversification and reducing carbon-intensive generation.
In addition to their differing energy profiles, all three countries possess substantial freshwater resources, favorable solar conditions, and growing interest in renewable-energy development. These characteristics make them suitable candidates for assessing the technical potential and deployment opportunities of FSPV systems under different energy-system conditions.

     \begin{figure*}[!htbp]
    \centering
    \includegraphics[width=\textwidth]{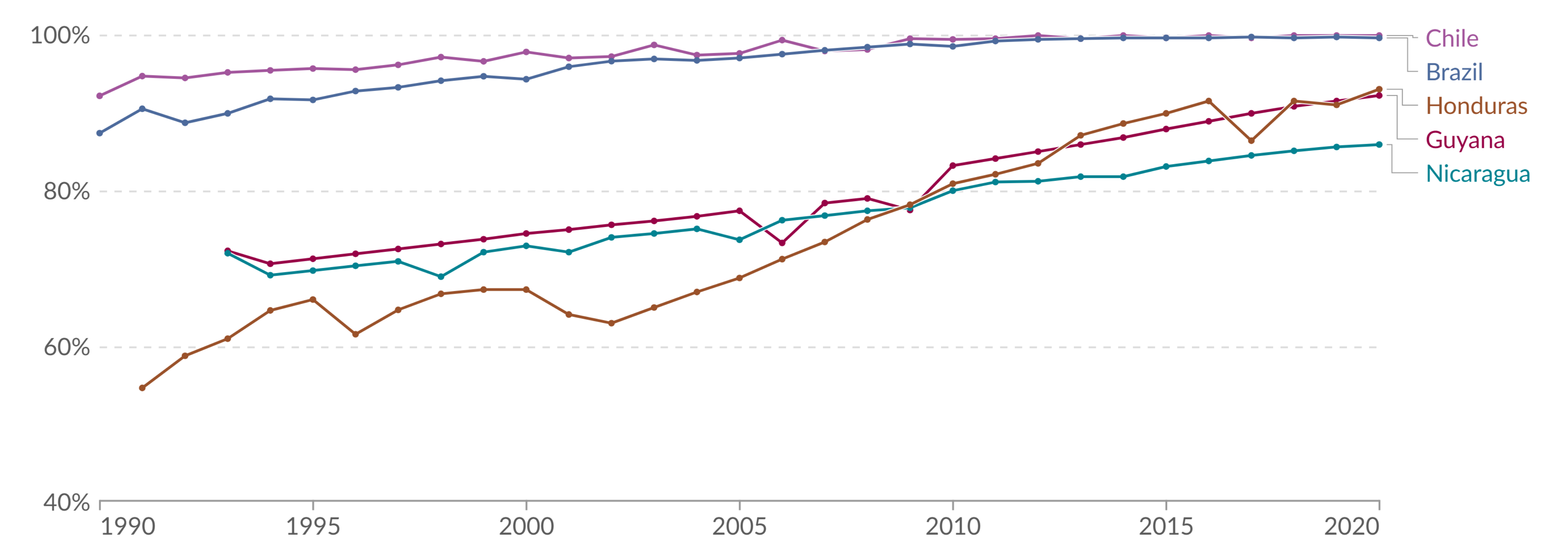}
    \caption{Share of the population with access to electricity for Nicaragua, Honduras, and Guyana (2020). Data source: \cite{ourworldindata_0} }
    \label{fig:fivecountrylinechart}
\end{figure*}

\subsection{Selection of representative water bodies}

\begin{figure*}[!htbp]
    \centering
    \includegraphics[width=0.55\textwidth]{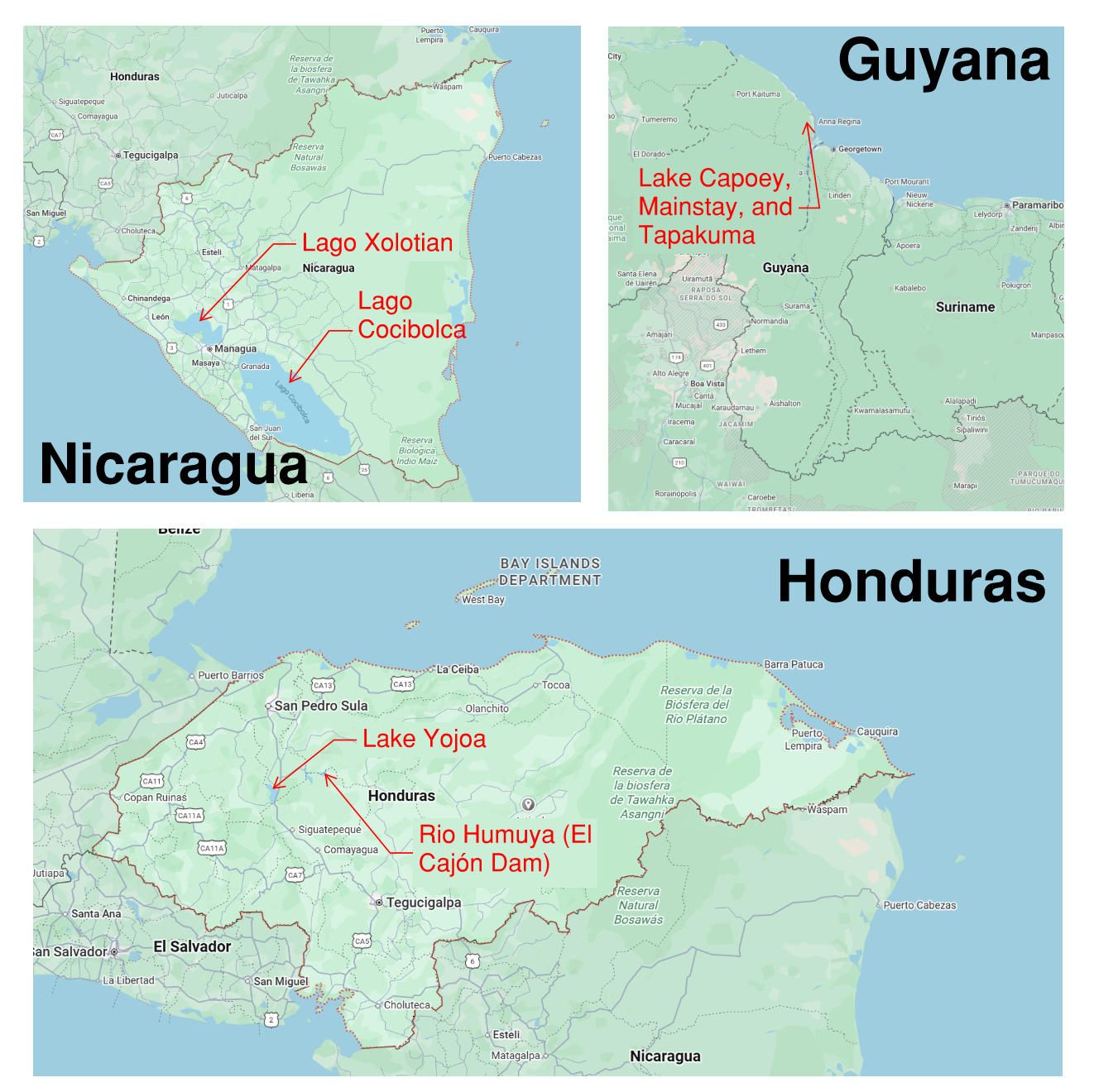}
    \caption{Waterbodies of interest for FSPV deployment and analysis in Nicaragua, Honduras, and Guyana.}
    \label{fig:waterbodiesofinterst}
\end{figure*}

Following the selection of Nicaragua, Honduras, and Guyana, representative water bodies were identified based on their surface area, accessibility, renewable-energy integration potential, and suitability for large-scale floating solar deployment. Figure \ref{fig:waterbodiesofinterst} presents the geographical distribution of the selected sites, including Lake Xolotlán and Lake Cocibolca in Nicaragua, Lake Yojoa and El Cajón Reservoir in Honduras, and the Capoey–Mainstay–Tapakuma water system in Guyana. \\
The selected sites represent diverse deployment conditions, including natural lakes, hydropower reservoirs, and interconnected freshwater systems. Lake Xolotlán and Lake Cocibolca provide extensive water surfaces suitable for utility-scale FSPV deployment, while Lake Yojoa and El Cajón Reservoir offer opportunities to investigate the integration of floating solar with existing hydropower infrastructure. In Guyana, the Capoey–Mainstay–Tapakuma system was selected due to its strategic importance for regional water-resource management and its potential to support future renewable-energy development.
Collectively, these sites capture a range of hydrological, geographical, and energy-system characteristics, enabling a comprehensive assessment of FSPV deployment opportunities across different operating environments. The selected water bodies therefore provide a suitable basis for evaluating technical performance, energy-generation potential, and renewable-energy integration pathways within the study region.

\section{FSPV system configuration and performance analysis}
This section presents the technical configuration and performance assessment of representative Floating Solar Photovoltaic (FSPV) systems deployed across selected waterbodies in Nicaragua, Honduras, and Guyana. Based on the resource assessment and deployment potential summarized in Table \ref{tab:fspv_potential}, representative FSPV capacities ranging from 50 MW to 398 MW are selected to evaluate system performance under site-specific climatic and geographical conditions. The analysis includes the electrical system architecture, collector substation design, cable configuration, and detailed energy-performance evaluation using key indicators such as annual energy generation, energy yield, capacity factor, and performance ratio. The selected case studies provide a comparative assessment of FSPV deployment across natural lakes and hydropower reservoirs while highlighting their potential contribution to renewable-energy expansion in the region.

\subsection{FSPV system configuration for PV array, inverter, collector substation, and cables} 

\subsubsection{The electrical system layout}
\begin{figure*}[!htbp]
    \centering
    \includegraphics[width=\textwidth]{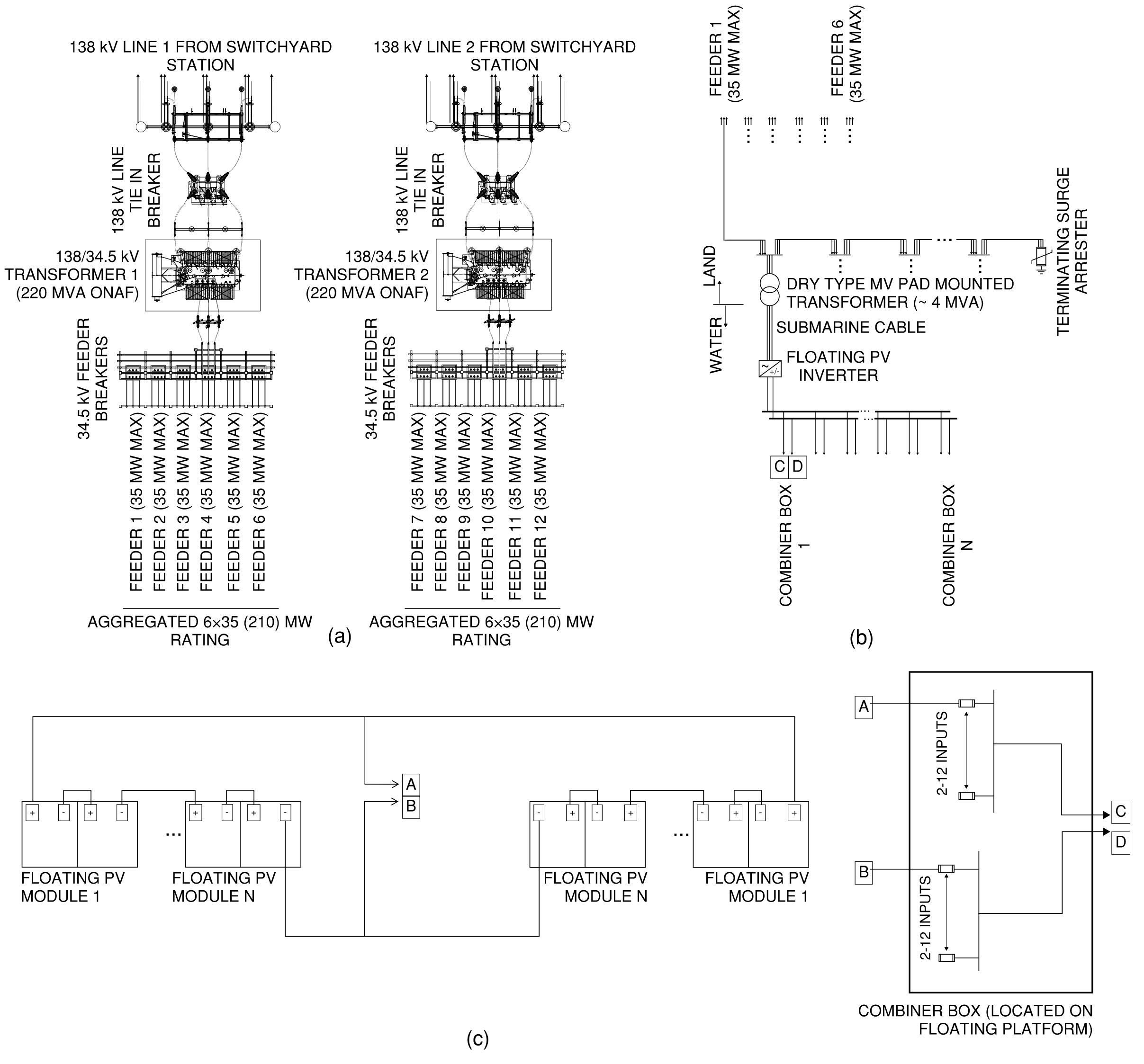}
    \caption{138/34.5 kV collector and lower level electrical architecture. (a) The collector substation with 138/34.5 kV generation step-up transformers. (b) Layout of the 34.5 kV feeder system, dry-type MV transformer, and floating PV inverter. (c) The module-level architecture with modules connected to combiner boxes.}
\label{fig:FSPV_Substation_and_Lower_Level_Electrical_Architecture}
\end{figure*}

For a grid-connected FSPV system, the collector substation is an essential part of a floating photovoltaic (FPV) power system. It serves as the main connection point between the solar arrays on the water and the wider electrical grid. Because of its importance, both the substation’s internal design and the downstream electrical layout need careful discussion. These collector substations are typically located on land and connect to the utility transmission network at voltages of 230 kV, 138 kV, or 69 kV. As shown in Figure \ref{fig:FSPV_Substation_and_Lower_Level_Electrical_Architecture} (a), a typical collector substation includes a main power transformer that steps up the generated voltage for transmission. These main power transformers usually has a capacity between 50 and 250 MVA and is built as a two‑winding unit, with the high‑voltage side in a delta configuration and the low‑voltage side in a resistance‑grounded wye configuration. The low‑voltage winding is normally rated at 34.5 kV and can carry a maximum ampacity rating between 4000 and 5000 A.

At the 34.5 kV level, the incoming feeders are terminated. These feeders can originate solely from the floating solar arrays or from a combination of floating solar and hydropower generation in co‑located hydro‑FPV (floating solar photovoltaic) systems. As shown in Figure \ref{fig:FSPV_Substation_and_Lower_Level_Electrical_Architecture} (b), the feeders are connected through a series of land-based sectionalizing cabinets. Each cabinet receives power from a land‑based dry‑type transformer, which is linked via a submarine cable to a PV inverter located on the floating PV platform. These dry‑type transformers are usually rated at 3-4 MVA and are preferred over oil-based transformers because they mitigate concerns about oil leakage near sensitive marine ecosystems. Figure \ref{fig:FSPV_Substation_and_Lower_Level_Electrical_Architecture} (c) further illustrates how the power output from the floating PV modules is arranged in a series‑parallel configuration, then routed through combiner boxes and ultimately fed into the PV inverter.

\subsubsection{Other grid level considerations for FSPV collector substation} 




    Integrating large renewable plants into the grid requires careful attention to how they supply and absorb reactive power, how their main transformers are rated, and how their inverters are controlled. Additionally, for system sizing, developers must use inverter capability curves that reflect real site conditions such as high ambient temperatures and actual grid voltage at the point of interconnection. Studies must verify that the plant can meet grid-code power factor requirements at the grid connection point while avoiding overload of the main step‑up power transformer, once cable losses, transformer losses, and auxiliary loads are all included. In some cases, developers may need to re‑nameplate transformers or reduce contracted export capacity if the transformer would otherwise be overstressed for significant periods \cite{ghosh2025integration}.

    On the control side, options exist between the selection of grid‑following inverters, which “follow” an existing grid, and grid‑forming inverters, which can set and stabilize voltage and frequency themselves \cite{li2022revisiting}. Grid‑forming units are particularly valuable in weak or highly renewable grids, but they are not universally superior. A well‑tuned grid‑following plant can perform adequately in many systems \cite{ospina2025grid}. Finally, the importance of designing for harmonic performance must be emphasized such that the PV plant attenuates problematic harmonics when measurements are taken at the point of interconnection, helping the plant comply with standards such as IEEE 519 and IEEE 2800 while controlling cost.

\subsubsection{Cable system and best practices}

The cabling in an FSPV plant can be categorized into three clusters: a. the land-based feeder cables that interface with the submarine cables and bring in the power generated by the floating solar modules to the main power transformer located in the substation. The current limit of these land-based feeder cables is determined by the operating temperature rise of the conductor in contact with the insulation. The Neher-McGrath
method and IEC 60287 are standard methods for calculating underground cable ampacity (current-carrying capacity) based on heat dissipation \cite{ramirez2020cables}. b. the cabling at the floating platform is usually channeled through cable trays and specialized conduit systems, providing a protected water-sealed path. c. submarine cable connecting the floating FSPV plant to an onshore connection point. Traditionally, submarine cables have been used in undersea HVDC transmission and offshore wind applications, but have witnessed an expanded application in the FSPV domain. Submarine cables are larger and heavier than land-based cables. The burial method for submarine cables depends on the sediment type of the water body, with simultaneous laying and burial using a jet plow being the most common for very soft to soft sediment areas. When hard sediments and rock are encountered, pre-trench and backfill is the preferred installation method. For these submarine cables, joints are avoided, and the installation is made in one continuous run, often ranging between 1 and 12 miles. For designing and installing submarine cables for FSPV installations, IEEE 1120-2004 serves as an useful guide. 
 
\subsection{Performance analysis of FSPV at major waterbodies at Nicaragua, Honduras, and Guyana}

Table \ref{tab:fspv_potential} outlines estimates of the theoretical maximum generation limits for the waterbodies under consideration, which forms the basis for the detailed performance analysis carrier out in the subsequent sections: \S\ref{subsec_Lake Xolotlán_Nicaragua} - \S\ref{subsec_Lake Capoey, Mainstay, and Tapakuma - Guyana}. For extremely large to large water bodies, such as Lake Xolotian, Lake Cocibolca, and Lake Yojoa, we limit the demonstration of FSPV implementation to an initial 50 MWac for a theoretical comparison. For brevity, the performance analysis will highlight representative cases: a) a 50 MWac FSPV implementation for Lake Xolotian, Nicaragua, b) a 398 MWac implementation at the El Cajon dam in Honduras, and c) a 95 MWac implementation at Lake Capoey, Mainstay, and Tapakuma in Guyana. 
    
\begin{table}[t]
\small
\centering
\caption{Potential estimation for FSPV deployment over selected waterbodies in Nicaragua, Honduras, and Guyana}
\label{tab:fspv_potential}
\renewcommand{\arraystretch}{1.2} 
\begin{tabular}{p{2.5cm} p{2.25cm} p{2.5cm} p{3cm} p{4cm}}
\hline
\textbf{Name of waterbody} &
\textbf{Total size of water body (acres)} &
\textbf{Usable size of waterbody for FSPV (\% / acres)\textsuperscript{*}} &
\textbf{Theoretical max generation at 0.4 MWdc per acre (MWdc)} &
\textbf{Theoretical max generation (MWac) and demonstration MWac} \\
\hline
\multicolumn{5}{l}{\textbf{Nicaragua}} \\
Lake Xolotian &
256{,}000 &
5--10\% / 12{,}800--25{,}600 &
5{,}120--10{,}240 &
3{,}938--7{,}876 max / 50 MWac demonstration in section \S\ref{subsec_Lake Xolotlán_Nicaragua}\\
Lake Cocibolca &
2{,}042{,}880 &
5--10\% / 102{,}144--204{,}288 &
40{,}857--81{,}715 &
31{,}428--62{,}857 / 50 MWac;  demonstration comparable to that presented in \S\ref{subsec_Lake Xolotlán_Nicaragua}\\
\hline
\multicolumn{5}{l}{\textbf{Honduras}} \\
Lake Yojoa &
19{,}840 &
10--30\% / 1{,}984--5{,}952 &
793--2{,}380 &
610--1{,}831 / 50 MWac;  demonstration comparable to that presented in \S\ref{subsec_Lake Xolotlán_Nicaragua}\\
El Cajon dam (reservoir area, also see Figure \ref{fig:hydrodam_El_Cajon_Dam_}) &
2{,}589 &
50--70\% / 1{,}294--1{,}812 &
517--724 &
398--557 / 398 MWac demonstration in \S\ref{subsec_El Cajón Dam Reservoir - Honduras}\\
\hline
\multicolumn{5}{l}{\textbf{Guyana}} \\
Lake Capoey, Mainstay, and Tapakuma &
1{,}030 &
30\% / 309 &
123 &
95 / 95 MWac demonstration  in \S\ref{subsec_Lake Capoey, Mainstay, and Tapakuma - Guyana}\\
\hline
\end{tabular}
\\[3pt]
\footnotesize\textsuperscript{*}Usable size of
waterbody for FSPV depending on overall size of waterbody, usually higher for smaller waterbodies and for reservoir area behind a dam.
\end{table}

\subsubsection{Lake Xolotlán - Nicaragua}
\label{subsec_Lake Xolotlán_Nicaragua}
Lake Xolotlán is one of the largest freshwater bodies in Nicaragua and presents a strategically important site for FSPV deployment. The lake is located in the western part of the country, adjacent to the capital city Managua, which makes it highly accessible from both an infrastructure and grid connectivity perspective. Its proximity to a major urban load center is particularly advantageous, as it reduces transmission losses and facilitates easier integration of generated power into the national grid.

From a geographical standpoint, Lake Xolotlán extends over an approximate surface area of 1,040 km², making it one of the largest lakes in Nicaragua. The lake has an average depth of around 9–10 m, with several near-shore regions exhibiting even shallower depths. Such bathymetric characteristics are advantageous for floating PV deployment, as they simplify anchoring and mooring design and reduce installation complexity compared to deeper water bodies. In addition, the lake typically experiences low to moderate wave activity, particularly in sheltered zones, which supports structural stability and reduces mechanical stress on floating platforms. The surrounding topography is relatively flat, especially near the Managua region, further facilitating installation logistics and grid connectivity. These combined geographical and physical characteristics make Lake Xolotlán a technically viable and scalable site for floating solar PV systems.

As shown in Table \ref{tab:fspv_potential}, Lake Xolotlán has a total surface area of approximately 256,000 acres. Assuming 5–10\% utilization of the available water surface for FSPV deployment, the usable area ranges from 12,800 to 25,600 acres, corresponding to a theoretical generation potential of 3,938–7,876 MWac. For comparative analysis, a 50 MWac FSPV system is considered in this study. The climatic conditions of Lake Xolotlán are summarized in Table \ref{tab:lake_xolotlan}.

\begin{table}[t]
\small
\centering
\caption{Weather parameters of Lake Xolotlán}
\label{tab:lake_xolotlan}
\renewcommand{\arraystretch}{1.2} 
\begin{tabular}{p{3cm} p{2.5cm} p{2.5cm}}
\hline
\textbf{Parameter} &
\textbf{Value/range} &
\textbf{Description}\\
\hline
Climate type &
Tropical savanna (Aw) &
Distinct wet and dry seasons\\

Mean annual temperature &
26–28 °C &
Warm conditions with low annual variation\\

Annual rainfall &
900–1100 mm &
Strong seasonal distribution\\

Wet season &
May – November &
High cloud cover and precipitation\\

Dry season &
November – April &
Low rainfall and high solar availability\\

Relative humidity &
60–84\% &
Higher during wet season\\

Average wind speed &
12–26 km/h &
Moderate, with higher values in dry months\\

Solar resource (GHI) &
5–6 kWh/m²/day &
High solar energy potential\\

Water surface temperature &
28–29 °C &
Supports module cooling effect \\

\hline
\end{tabular}
\end{table}

The climatic characteristics of Lake Xolotlán, summarized in Table \ref{tab:lake_xolotlan}, indicate a favorable environment for FSPV deployment. The region experiences a tropical savanna climate with distinct wet and dry seasons, which directly influence solar availability. The dry season (November–April) supports higher irradiance and improved system output, while the wet season (May–November) introduces moderate reductions due to increased cloud cover. The mean annual temperature ranges from 26–28°C, while average wind speeds vary between 12 and 26 km/h. The site also benefits from a strong solar resource of approximately 5–6 kWh/m²/day and relatively stable water surface temperatures of 28–29°C, which support module cooling and improved system performance.

The solar resource profile indicates consistently high global horizontal irradiance (GHI) throughout the year. Monthly average GHI varies from 434.87 W/m² in June to 561.61 W/m² in March, with peak values occurring during the dry season when cloud cover is minimal. Although irradiance decreases during the rainy season, values remain above 430 W/m² throughout the year, indicating a stable solar resource. Such limited seasonal variability supports reliable and predictable power generation and contributes to the strong performance expected from the FSPV system. 

\begin{figure*}[!htbp]
    \centering
    \includegraphics[width=0.75\textwidth]{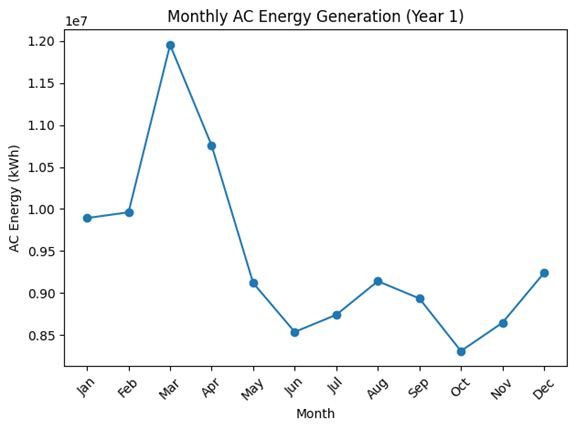}
    \caption{Monthly power generation of the 50 MW, size as established in Table \ref{tab:fspv_potential}, FSPV system at Lake Xolotian.\\
    }
    \label{fig:FSPV_system_Lake_Xolotian_}
\end{figure*}

Building on these favorable site characteristics, a detailed performance assessment was carried out for a 50 MW FSPV system. The monthly energy-generation profile is presented in Figure \ref{fig:FSPV_system_Lake_Xolotian_}. The system achieves its highest output in March (11.95 GWh), followed by April (10.76 GWh), corresponding to the peak dry season in Nicaragua when solar irradiance is highest. A decline in generation is observed from May through October, with the minimum output recorded in October (8.31 GWh), coinciding with increased cloud cover and rainfall during the wet season. Despite this seasonal reduction, monthly generation remains consistently above 8 GWh, demonstrating a stable solar resource and reliable year-round electricity production.

The yearly performance metrics of the FSPV system are summarized in Table \ref{tab:lake_xolotlan_metrics}. The system generates 113.24 GWh of AC electricity during the first year of operation. This corresponds to an energy yield of 2,265 kWh/kW and a capacity factor of 25.9\%, highlighting the strong solar resource available at the site. The performance ratio of 0.82 indicates efficient system operation and effective management of system losses. These results demonstrate the strong technical potential of FSPV deployment at Lake Xolotlán and confirm the suitability of the site for utility-scale implementation.

\begin{table}[t]
\small
\centering
\caption{Performance metrics of the 50 MW FSPV system at Lake Xolotlán}
\label{tab:lake_xolotlan_metrics}
\renewcommand{\arraystretch}{1.2} 
\begin{tabular}{p{4cm} p{3cm} }
\hline
\textbf{Metrics} &
\textbf{Value} \\
\hline
Annual AC energy in Year 1 &
113,238,480 kWh \\

DC capacity factor in Year 1 &
25.9\% \\

Energy yield in Year 1 &
2,265 kWh/kW \\

\hline
\end{tabular}
\end{table}

\begin{figure*}[!htbp]
    \centering
    \includegraphics[width=0.75\textwidth]{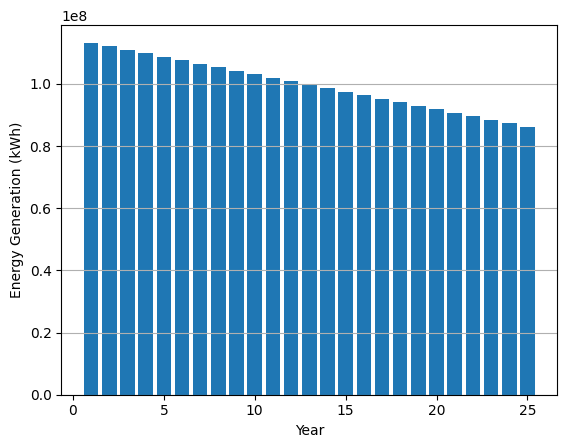}
    \caption{Life time energy generation of the 50 MW, size as established in Table \ref{tab:fspv_potential}, FSPV system at Lake Xolotian.}
    \label{fig:FSPV_system_Lake_Xolotian_lifetime}
\end{figure*}

The lifetime energy-generation profile of the 50 MW FSPV system is presented in Figure \ref{fig:FSPV_system_Lake_Xolotian_lifetime}. Assuming an annual module degradation rate of 1\%, annual electricity generation decreases gradually from 113.24 GWh in Year 1 to approximately 86.18 GWh by Year 25. The decline follows a predictable trend and reflects normal module aging. Despite this reduction, the system maintains substantial electricity production throughout its operational lifetime, generating more than 100 GWh annually for over a decade and remaining above 86 GWh even in the final year of operation.

The performance of the 50 MW land-based PV (LBPV) system was evaluated under the same climatic conditions as the FSPV system at Lake Xolotlán, enabling a direct comparison. The LBPV system delivers an annual AC energy output of 106.86 GWh in the first year, which is lower than the 113.24 GWh obtained from the FSPV system. This difference reflects the influence of operating conditions rather than differences in solar resource, as both systems are exposed to the same irradiance profile. A similar trend is observed in the energy yield, where the LBPV system achieves 2,137 kWh/kW compared to 2,265 kWh/kW for the FSPV system. The capacity factor of the LBPV system is 24.4\%, compared to 25.9\% for the FSPV system, while the performance ratio decreases from 0.82 to 0.78. These results indicate that the FSPV system outperforms the LBPV system across all key performance indicators. The improvement is primarily attributed to the cooling effect of the water surface, which reduces module operating temperatures and associated thermal losses. Consequently, the FSPV system provides higher annual energy generation and improved operational performance, highlighting its suitability for large-scale deployment at Lake Xolotlán.

\subsubsection{El Cajón Dam Reservoir - Honduras}
\label{subsec_El Cajón Dam Reservoir - Honduras}
\begin{figure*}[!htbp]
    \centering
    \includegraphics[width=0.75\textwidth]{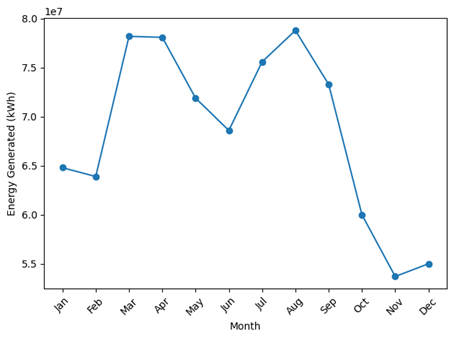}
    \caption{Monthly power generation of the 398 MW, size as established in Table \ref{tab:fspv_potential}, FSPV system at El Cajón Reservoir.}
    \label{fig:FSPV_system_lake_Elcajon}
\end{figure*}

The El Cajón Dam reservoir, formed along the Humuya River in Honduras, is selected for large-scale FSPV deployment due to its extensive water surface, existing hydropower infrastructure, and strategic grid connectivity. As one of the largest reservoirs in the region, it offers significant potential for utility-scale installation without competing for land resources. The El Cajón Reservoir was selected for a 398 MW FSPV demonstration system because Table \ref{tab:fspv_potential} indicates a theoretical deployment potential of 398–557 MW based on 50–70\% utilization of the available reservoir area. The lower bound of 398 MW was adopted as a conservative and technically feasible scenario while still representing a utility-scale hydro–solar co-located installation. The presence of established transmission infrastructure associated with the dam further reduces integration challenges and enhances the feasibility of deploying a high-capacity (398 MW) system. In addition, the reservoir experiences favorable solar resource conditions typical of tropical regions, supporting high energy generation potential. From a bathymetric perspective, the reservoir is characterized by a large surface area of approximately 94 km² and a maximum depth exceeding 90 m, with significant spatial variation due to its riverine origin. While deeper zones exist near the dam structure, a substantial portion of the reservoir consists of moderate-depth regions suitable for FSPV deployment. These areas provide flexibility in system placement and allow for optimized anchoring strategies using mooring systems adapted to variable depths.\\
The reservoir also experiences seasonal water level fluctuations, driven by hydropower operations and inflow variability. Such fluctuations can be significant but are generally predictable, allowing for the design of adaptive mooring systems to maintain platform stability. Overall, the combination of large surface area, existing electrical infrastructure, and manageable bathymetric conditions makes the El Cajón reservoir a technically viable and strategically important site for large-scale FSPV implementation. \\
The solar resource at El Cajón Dam reservoir exhibits moderate to high global horizontal irradiance (GHI) with clear seasonal variation. The monthly average GHI ranges from 367.55 W/m² (December) to 540.81 W/m² (April), with peak values occurring during the late dry season (March–April). This period is characterized by reduced cloud cover and enhanced solar availability, which is expected to support higher energy generation from the FSPV system. A slight decline in irradiance is observed during May and June with the onset of the wet season, although the reduction remains moderate compared to other tropical locations. Interestingly, irradiance levels recover and remain relatively high during July and August, reaching values above 520 W/m², indicating intermittent clear-sky conditions even within the wet season. This contributes to a more distributed generation profile across the year. From September onwards, the GHI gradually decreases, reaching lower values in November and December, reflecting increased cloud cover and seasonal transition. Despite this variation, the overall irradiance remains within a favorable range throughout the year.  The monthly energy generation of El Cajon dam reservoir is given in Fig. \ref{fig:FSPV_system_lake_Elcajon}.

The monthly energy generation profile of the 398 MW FSPV system at El Cajón Dam reservoir reflects the relatively balanced solar resource identified in the preceding analysis. The system achieves high generation during March and April ($\approx$ 78 GWh), corresponding to peak irradiance in the late dry season. This high-output period extends into August, where the system records the maximum generation of approximately 79 GWh, supported by sustained irradiance levels even during the mid-year months. \\
Unlike smaller inland lakes, the decline in generation during the wet season is less pronounced. Although a reduction is observed in June ($\approx$ 68.5 GWh), the system quickly recovers in July and August, maintaining generation above 75 GWh. This behavior aligns with the observed GHI pattern, where intermittent clear-sky conditions during the wet season help sustain energy production.

A sharper decline is observed from October onwards, with generation dropping to around 60 GWh in October and reaching a minimum of approximately 53–55 GWh in November and December. This trend corresponds to reduced irradiance and increased cloud cover during the late wet season and transition period. Despite this drop, the overall variation remains relatively moderate considering the large system scale. The yearly performance of the system is given in Table \ref{tab:lake_El_Cajón_Dam_reservoir_metrics}. The performance of the 398 MW FSPV system at El Cajón Dam reservoir reflects strong and stable energy generation, consistent with the favorable solar resource identified earlier. The system produces 821.74 GWh in the first year, indicating high energy output at utility scale and effective utilization of the available irradiance.

\begin{table}[t]
\small
\centering
\caption{Performance metrics of the 398 MW FSPV system at El Cajón Dam reservoir}
\label{tab:lake_El_Cajón_Dam_reservoir_metrics}
\renewcommand{\arraystretch}{1.2} 
\begin{tabular}{p{4cm} p{3cm} }
\hline
\textbf{Metrics} &
\textbf{Value} \\
\hline
Annual AC energy in Year 1 &
821,744,960 kWh\\

DC capacity factor in Year 1 &
23.6\% \\

Energy yield in Year 1 &
2,065 kWh/kW\\

Performance Ratio (Year 1) &
0.81\\

\hline
\end{tabular}
\end{table}

The energy yield of 2,065 kWh/kW is relatively high, confirming efficient conversion of solar energy into electrical output. This is supported by the DC capacity factor of 23.6\%, which indicates sustained generation across the year with limited seasonal disruption. The balanced irradiance profile, including strong mid-year solar availability, contributes to maintaining this level of performance.

The performance ratio of 0.81 suggests well-controlled system losses and efficient operation. Despite the large system size, the relatively high PR indicates that thermal and operational losses are effectively managed. The cooling effect of the reservoir likely plays a key role in maintaining module efficiency, particularly under warm climatic conditions.

\begin{figure*}[!htbp]
    \centering
    \includegraphics[width=0.75\textwidth]{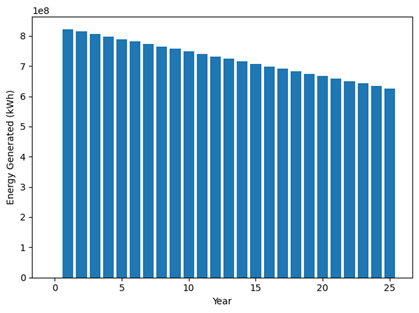}
    \caption{Life time energy generation of the 398 MW, size as established in Table \ref{tab:fspv_potential}, FSPV system at El Cajón Reservoir.}
    \label{fig:FSPV_system_lake_Elcajon_lifetime}
\end{figure*}

The long-term energy generation profile of the 398 MW FSPV system at El Cajón Dam reservoir presented in Fig \ref{fig:FSPV_system_lake_Elcajon_lifetime} shows a steady and predictable decline over the 25-year operational period, consistent with the assumed module degradation. The system generates approximately 821.7 GWh in the first year, which gradually decreases to around 630–640 GWh by year 25, representing an overall reduction of nearly 23–24\%.

The decline follows a smooth and nearly linear trend, indicating stable system operation without abrupt losses or performance irregularities. Despite the reduction, the system maintains a high level of generation throughout its lifetime, consistently exceeding 700 GWh for more than half of the operational period. This sustained output highlights the advantage of deploying large-scale FSPV systems in reservoirs with strong and balanced solar resources.

When considered alongside the monthly and annual performance results, the degradation trend confirms that the primary driver of long-term variation is module aging rather than environmental instability. The relatively high initial energy yield and capacity factor help offset degradation effects, ensuring robust lifetime energy production.

\subsubsection{Lake Capoey, Mainstay, and Tapakuma - Guyana}
\label{subsec_Lake Capoey, Mainstay, and Tapakuma - Guyana}
Lake Capoey, Mainstay, and Tapakuma are selected as an additional study site to evaluate FSPV performance under the tropical climatic conditions of Suriname. The lake offers sufficient water surface area for utility-scale deployment without creating competition for agricultural or urban land. Its location in a region with high solar availability and relatively stable temperatures makes it suitable for reliable year-round energy generation. From a bathymetric perspective, the lake is characterized by shallow to moderate water depths, which are favorable for anchoring and mooring design. Most regions are expected to fall within manageable depth ranges, reducing installation complexity and structural requirements. In addition, the lake experiences limited water level variation, which supports stable platform operation and minimizes mechanical stress on the floating structure. These geographical and hydrological conditions provide a suitable basis for FSPV deployment. Therefore, a 95 MW FSPV system is designed at Lake Capoey, Mainstay, and Tapakuma to evaluate its technical performance under site-specific climatic and operational conditions.
The solar resource at Lake Capoey, Mainstay, and Tapakuma is characterized by consistently high global horizontal irradiance (GHI) throughout the year, indicating strong potential for FSPV deployment. The monthly average GHI ranges from 401.75 W/m² (December) to 544.56 W/m² (September), with relatively small variation across the year compared to other tropical sites.

Unlike locations where peak irradiance occurs during the dry season, Lake Capoey, Mainstay, and Tapakuma cluster exhibits strong solar availability during both mid-year and late-year periods. Irradiance remains above 440 W/m² for most of the year, with particularly high values observed from July to November. The maximum GHI in September suggests that the site experiences sustained solar availability even during periods that may otherwise coincide with higher rainfall.
The relatively narrow spread between minimum and maximum irradiance indicates low seasonal variability, which is advantageous for maintaining stable energy generation throughout the year. Compared to sites with more pronounced wet-season reductions, the solar resource at Lake Capoey, Mainstay, and Tapakuma is more evenly distributed, suggesting that the FSPV system may experience less fluctuation in monthly output.

\begin{figure*}[!htbp]
    \centering
    \includegraphics[width=0.75\textwidth]{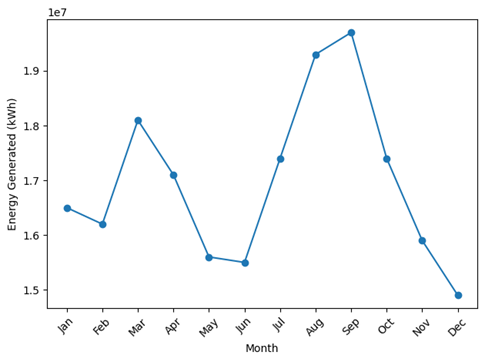}
    
    \caption{Monthly power generation of the 95 MW, size as established in Table \ref{tab:fspv_potential}, FSPV system at Lake Capoey, Mainstay, and Tapakuma.}
    \label{fig:FSPV_system_lake_Capoey}
\end{figure*}

The monthly energy generation profile of the 95 MW FSPV system at Lake Capoey, Mainstay, and Tapakuma, shown in Fiure \ref{fig:FSPV_system_lake_Capoey}, reflects the relatively uniform irradiance pattern observed at the site. Monthly generation ranges from approximately 14.9 GWh (December) to 19.6 GWh (September), indicating moderate seasonal variation compared to other tropical locations.
Higher generation is observed during March and from August to October, with the peak output occurring in September. This trend closely follows the GHI profile, where irradiance remains elevated during the second half of the year. In particular, the strong generation during August and September suggests that the site benefits from sustained solar availability even during periods that may typically experience increased rainfall.

Lower output is recorded during May, June, and December, with generation remaining between 15–16 GWh. However, the decline is relatively modest, and the system does not experience the sharp seasonal reductions observed at some of the previously studied sites. This indicates that the solar resource at Lake Capoey, Mainstay, and Tapakuma is more evenly distributed across the year, resulting in a smoother monthly generation profile. Table \ref{tab:lake_Capoey_metrics} presents the performance metrices of the 95 MW FSPV system.

\begin{table}[t]
\small
\centering
\caption{Performance metrices of the 95 MW FSPV system at Lake Capoey, Mainstay, and Tapakuma}
\label{tab:lake_Capoey_metrics}
\renewcommand{\arraystretch}{1.2} 
\begin{tabular}{p{4cm} p{3cm} }
\hline
\textbf{Metrics} &
\textbf{Value} \\
\hline
Annual AC energy in Year 1 &
203,708,624 kWh\\

DC capacity factor in Year 1 &
24.5\% \\

Energy yield in Year 1 &
2,144 kWh/kW\\

Performance Ratio (Year 1) &
0.83\\

\hline
\end{tabular}
\end{table}

\begin{figure*}[!htbp]
    \centering
    \includegraphics[width=0.75\textwidth]{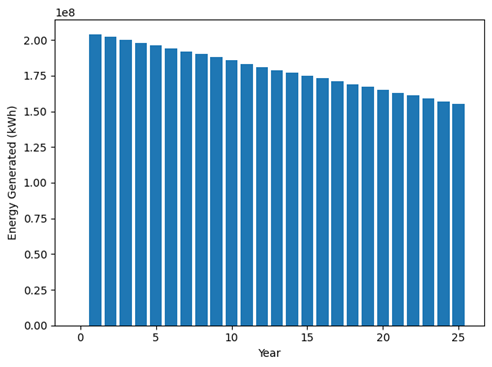}
    \caption{Life time energy generation of the 95 MW, size as established in Table \ref{tab:fspv_potential}, FSPV system at Lake Capoey, Mainstay, and Tapakuma cluster.}
    \label{fig:FSPV_system_lake_Capoey_lifetime}
\end{figure*}

The long-term energy generation profile of the 95 MW FSPV system at Lake Capoey, Mainstay, and Tapakuma, as seen in Figure \ref{fig:FSPV_system_lake_Capoey_lifetime}, shows a gradual and predictable decline over the 25-year operational period, consistent with the assumed annual module degradation. The system generates approximately 204 GWh in the first year, which decreases steadily to around 156 GWh by year 25, in a linear manner, indicating stable operation without abrupt losses or performance anomalies. Despite the reduction, the system continues to maintain strong annual generation throughout its lifetime, remaining above 175 GWh for more than half of the project duration. This sustained performance reflects the stable solar resource and relatively low seasonal variability observed at the site.
The performance of the 95 MW land-based PV (LBPV) system at Lake Capoey–Mainstay–Tapakuma is lower than that of the FSPV configuration under the same site conditions. The LBPV system generates 192.92 GWh in the first year, compared to approximately 204 GWh for the FSPV system, corresponding to a reduction of around 5–6\% in annual energy output. A similar trend is observed in the energy yield, where the LBPV system achieves 2,031 kWh/kW, which is lower than the yield obtained from the FSPV system. The DC capacity factor of 23.2\% also indicates slightly lower utilization of the available solar resource. In addition, the performance ratio of 0.78 suggests marginally higher system losses in the land-based configuration.
These differences are primarily associated with higher module operating temperatures and greater soiling in land-based systems. In contrast, the FSPV system benefits from the cooling effect of the water surface, which helps maintain module efficiency and reduces thermal losses. This advantage becomes particularly important in tropical climates such as Guyana, where ambient temperatures remain high throughout the year.
Although the performance gap is moderate, the higher energy generation from the FSPV system can lead to meaningful gains in long-term electricity production. At utility scale, even a 5–6\% increase in annual generation can significantly improve project revenues and overall system utilization.

\section{Suitability and techno-economic analysis of FSPV for marginalized and undeserved areas}

\subsection{FSPV as an enabler in energy adoption in marginalized and undeserved areas}
Floating solar photovoltaic (FSPV) systems provide a viable approach to addressing land-use constraints in solar energy deployment. Conventional ground-mounted solar PV plants typically require between 3 and 5 acres of land per megawatt (MW) of installed capacity, translating to an average generation density of approximately 0.2 to 0.325 MWac per acre. In the case of utility-scale developments, the financial feasibility and long-term sustainability of these systems depend on parameters such as capital cost, levelized cost of energy (LCOE), and payback period \cite{allouhi2019energetic, gurturk2019economic}. In marginalized and underserved communities, however, land availability and affordability often present significant barriers to solar PV implementation. Pressure on agricultural land usage, urban growth, and rising land values directly influence the payback period and the LCOE, making conventional solar farms less economically attractive. Floating solar PV installations offer a promising alternative by utilizing existing water bodies, such as reservoirs, lakes, or irrigation canals, for energy generation. This approach eliminates the need for land acquisition and can substantially reduce capital expenditure.

Beyond land savings, FSPV systems provide several additional benefits for these communities. The presence of PV modules on the water surface reduces evaporation rates, thereby conserving water resources, which is critical in regions experiencing water scarcity. The cooling effect of the water also enhances module efficiency, typically resulting in 10 to 15 percent higher energy output compared to conventional land-based systems. These combined advantages make FSPV a technically and economically compelling option for expanding clean energy access in marginalized and underserved regions.

From a commercial site engineering standpoint, floating PV systems help overcome many of the challenges and project risks commonly faced in land-based solar projects. Land-based PV installations require extensive pre-development work, such as erosion control measures using silt fences, straw wattles, or compost filter socks \cite{sandlererosion}, to manage disturbances from grading and trenching activities. After installation, additional effort is needed to handle stormwater runoff from impervious areas like access roads, often through swales or drainage systems into detention ponds. These activities add cost, labor, and regulatory compliance risks. By contrast, floating PV largely avoids soil disturbance, grading, and erosion management needs since it is installed on water surfaces, significantly reducing both civil work complexity and related environmental risks.

\subsection{FSPV techno-economic analysis}

The economic feasibility of the proposed 50 MW floating solar photovoltaic (FSPV) system on Lake Managua, Nicaragua, was evaluated to assess its long-term financial viability and investment potential. Economic analysis plays an important role in determining whether large-scale FSPV deployment can provide cost-effective and sustainable electricity generation under regional climatic and operational conditions. In this study, key economic indicators such as capital cost, operation and maintenance (O\&M) cost, net present value (NPV), internal rate of return (IRR), payback period, and levelized cost of energy (LCOE) were considered for comprehensive financial assessment.
Lake Managua was selected due to its large available water surface area, favorable solar resource availability, and proximity to major electricity demand centers. The utilization of floating PV technology on the lake can reduce land acquisition requirements while improving module performance through the natural cooling effect of water. These advantages can enhance annual energy generation and contribute to improved economic returns over the project lifetime.
Since all the selected case study lakes considered in this work were designed with the same installed capacity of 50 MW, the economic framework and major cost components remain nearly identical for the other lakes as well. Therefore, the detailed economic analysis is first presented for the Lake Managua FSPV system, while similar economic trends can be expected for the remaining lakes, with only slight variations due to differences in climatic conditions, energy generation, and site-specific operational parameters.
The economic assessment was carried out based on the simulated annual energy output of the 50 MW FSPV system. All major project expenses, including PV modules, floating structures, anchoring systems, inverters, electrical infrastructure, installation, and maintenance costs, were incorporated into the analysis. The obtained results provide insight into the commercial viability and future deployment potential of floating solar technology in Nicaragua.
The economic performance of the proposed 50 MW floating solar photovoltaic (FSPV) system was evaluated using several standard financial indicators, including the Net Present Value (NPV), Levelized Cost of Energy (LCOE), Internal Rate of Return (IRR), and payback period. These parameters provide a comprehensive understanding of the long-term economic viability and profitability of the project.
The Net Present Value (NPV) represents the difference between the present value of total benefits and the present value of total project costs over the system lifetime. A positive NPV indicates that the project is economically feasible and financially profitable. The NPV is calculated using Equation (\ref{eqn_NPV}) \cite{abdelhady2021performance}:


\begin{equation}
\label{eqn_NPV}
NPV = \sum_{t=0}^{N} \frac{C_t}{1 +r^{t}} \
\end{equation}

where \(C_t\) represents the net cash flow in year \(t\), \(r\) is the discount rate, and \(N\) is the project lifetime in years. The Levelized Cost of Energy \((LCOE)\) is one of the most important economic indicators for renewable energy systems. It represents the average cost of electricity generation over the entire project lifetime and is expressed in USD/kWh. Lower LCOE values indicate better economic performance. The LCOE is calculated using Equation (\ref{eqn_LCOE}) \cite{wang2025techno}:

\begin{equation}
\label{eqn_LCOE}
LCOE =  \frac{\sum_{t=0}^{N}\frac{I_t+O_t+M_t}{(1+r)^t}}{\sum_{t=0}^{N} \frac{E_t}{(1+r)^t}} \
\end{equation}

where \(I_t\) is the investment cost, \(O_t\) is the operation cost, \(M_t\) is the maintenance cost, and \(E_t\) is the energy generated during year \(t\).

The payback period (\(PBP\)) represents the time required to recover the initial investment through the annual revenue generated by the system. It is calculated using Equation (\ref{eqn_PBP}) \cite{abdelhady2021performance, wang2025techno}:

\begin{equation}
\label{eqn_PBP}
PBP =  \frac{\mbox{Initial Investment} }{\mbox{Annual Net Cash Flow}}
\end{equation}

The economic feasibility of the proposed 50 MW floating photovoltaic (FPV) system on Lake Managua was evaluated based on the total installed capacity and the projected annual energy generation obtained from the simulation results. FPV systems generally require higher initial investment costs compared to conventional land-based PV systems due to additional floating platforms, anchoring systems, and water-resistant electrical infrastructure. According to recent literature, the average capital expenditure (CAPEX) of large-scale FPV systems typically ranges between 1200 and 1800 USD/kW, depending on site-specific conditions and system configuration.

For the present study, an average capital cost of 1500 USD/kW was considered for the 50 MW FPV system installed on Lake Managua \cite{serat2025comparative, world2018sun}. Based on this assumption, the total initial investment cost of the proposed system was estimated to be approximately 75 million USD. Among the total CAPEX, floating structures and pontoon systems constitute one of the major additional expenses in FPV projects and generally account for nearly 20–30\% of the total investment cost. Therefore, the cost associated with floating platforms and floaters for the proposed Lake Managua FPV system was estimated to be approximately 15–22.5 million USD. Anchoring and mooring systems generally contribute an additional 5–10\% of the total CAPEX, corresponding to nearly 3.75–7.5 million USD for the proposed installation \cite{manolache2025floating, filgueira2025economic}. The annual operation and maintenance (O\&M) cost of FPV systems is generally considered to be around 1\%–2\% of the total capital investment. In this work, the annual O\&M cost was assumed to be 1.5\% of the total investment cost, corresponding to approximately 1.125 million USD per year.\\
The average electricity tariff considered for the economic analysis was based on the official industrial electricity rate in Nicaragua, which is approximately 0.212 USD/kWh. Using the annual energy generation obtained from the simulation results (113.24 million kWh in Year 1), the proposed FPV system is expected to generate approximately 24.0 million USD in annual revenue during the first year of operation. This estimation represents a realistic economic scenario and does not account for future increases in electricity prices, renewable energy incentives, or carbon credit benefits, which may further improve the financial attractiveness of the project.
Based on the estimated annual revenue and annual O\&M expenses, the net annual profit of the FPV system was determined for further financial assessment. Subsequently, the Levelized Cost of Energy (LCOE), Net Present Value (NPV), Internal Rate of Return (IRR), and discounted payback period were evaluated to determine the long-term economic viability of the proposed FPV installation on Lake Managua. Table \ref{tab:50MWFSPV} summarizes the key financial metrics for the proposed 50 MW FSPV system. The results show an LCOE of 0.0476 USD/kWh, an NPV of 67.8 million USD, an IRR of 24.1\%, and a discounted payback period of 5.1 years.

\begin{table}[t]
\small
\centering
\caption{50 MW FSPV system – Financial results}
\label{tab:50MWFSPV}
\renewcommand{\arraystretch}{1.2} 
\begin{tabular}{p{6cm} p{3cm} }
\hline
\textbf{Parameter} &
\textbf{Value} \\
\hline
Installed capacity (MW) &
50 \\

Total CAPEX (million USD) &
\$75.00 \\

Annual O\&M (million USD) &
\$1.125\\

Annual revenue (Year 1) (million USD) &
\$24.01\\

LCOE (USD/kWh) &
\$0.0476\\

NPV (million USD) &
\$67.8\\

IRR (\%) &
24.1\%\\

Discounted payback period (years) &
5.1\\

\hline
\end{tabular}
\end{table}

\subsubsection{Techno-economic analysis - Comparison of FSPV against a comparable 50 MW LBPV system}

The economic feasibility of the proposed 50 MW land-based photovoltaic (LBPV) system in the Lake Managua region was evaluated based on the total installed capacity and the projected annual energy generation obtained from the simulation results. Compared to floating photovoltaic systems, land-based PV systems generally require lower initial investment costs due to the absence of floating platforms, anchoring systems, and specialized water-resistant electrical infrastructure. However, additional expenses associated with land acquisition, site preparation, grading, and civil works must be considered in the overall project cost.

According to recent literature, the average capital expenditure (CAPEX) of utility-scale LBPV systems typically ranges between 800 and 1200 USD/kW depending on project scale, terrain conditions, grid connectivity, and regional installation costs \cite{kabeyi2023levelized, thai2023comparative}. For the present study, a capital cost of 1324 USD/kW was considered for the proposed 50 MW LBPV system in the Lake Managua region, accounting for site-specific costs including land development and grid interconnection infrastructure. Based on this assumption, the total initial investment cost of the proposed system was estimated to be approximately 66.2 million USD.

Land acquisition constitutes one of the major additional costs associated with utility-scale LBPV installations. In this study, a land requirement of approximately 125 hectares was considered based on an estimated land utilization factor of 2.5 hectares per MW \cite{valera2022deployment, joshi2026harnessing}. Accordingly, the land acquisition cost was estimated to be approximately 2.25 million USD based on prevailing industrial land prices in the region. The annual operation and maintenance (O\&M) cost of utility-scale LBPV systems is generally considered to be around 0.5\%–1.5\% of the total capital investment. In this work, the annual O\&M cost was assumed to be 1.0\% of the total investment cost, corresponding to approximately 0.66 million USD per year.

The average electricity tariff considered for the economic analysis was based on the industrial electricity tariff in Nicaragua, estimated at approximately 0.212 USD/kWh. Using the annual energy generation obtained from the simulation results (106.86 million kWh in Year 1), the proposed LBPV system is expected to generate approximately 22.65 million USD in annual revenue during the first year of operation. This estimation represents a practical economic scenario and does not include possible future increases in electricity tariffs, renewable energy incentives, or carbon credit benefits, which may further enhance the economic attractiveness of the project.

For the financial assessment, a 70:30 debt-to-equity financing structure with an 8\% loan interest rate over a repayment period of 15 years was considered. Based on these assumptions, the Levelized Cost of Energy (LCOE), Net Present Value (NPV), Internal Rate of Return (IRR), and discounted payback period were evaluated to determine the long-term economic viability of the proposed LBPV system in the Lake Managua region.
Table \ref{tab:50MWLBPV} summarizes the key financial metrics for the proposed 50 MW LBPV system. The results show an LCOE of 0.0403 USD/kWh, an NPV of 59.6 million USD, an IRR of 22.3\%, and a discounted payback period of 5.4 years.

\begin{table}[t]
\small
\centering
\caption{50 MW LBPV System – Financial Results}
\label{tab:50MWLBPV}
\renewcommand{\arraystretch}{1.2} 
\begin{tabular}{p{6cm} p{3cm} }
\hline
\textbf{Parameter} &
\textbf{Value} \\
\hline
Installed capacity (MW) &
50 \\

Total CAPEX (million USD) &
\$66.20 \\

Annual O\&M (million USD) &
\$0.45\\

Annual revenue (Year 1) (million USD) &
\$22.65\\

LCOE (USD/kWh) &
\$0.0403\\

NPV (million USD) &
\$59.6\\

IRR (\%) &
22.3\%\\

Discounted payback period (years) &
5.4\\

\hline
\end{tabular}
\end{table}

The comparative economic analysis between the proposed 50 MW floating solar photovoltaic (FSPV) system and the 50 MW land-based photovoltaic (LBPV) system in the Lake Managua region demonstrates that both configurations are economically feasible under the considered operating and financial conditions. However, notable differences were observed in terms of capital investment, energy generation, profitability, and long-term economic performance.
The FSPV system required a higher total capital investment of approximately 75 million USD (1,500 USD/kW) compared to 66.2 million USD (1,324 USD/kW) for the LBPV system, representing a 13.3\% higher upfront cost for the floating configuration. The higher CAPEX of the floating system is primarily associated with the additional costs of floating platforms and pontoon structures (accounting for 20–30\% of total CAPEX, approximately 15–22.5 million USD), anchoring and mooring systems (5–10\% of total CAPEX, approximately 3.75–7.5 million USD), and water-resistant electrical infrastructure. For the LBPV system, land acquisition costs contributed approximately 2.25 million USD, assuming a land requirement of 125 hectares (2.5 hectares per MW) at estimated local industrial land rates. Consequently, the annual operation and maintenance (O\&M) cost of the FSPV system was also higher, estimated at 1.125 million USD/year (1.5\% of CAPEX), whereas the LBPV system required only 0.45 million USD/year (1.0\% of CAPEX). The lower O\&M requirement for the LBPV system reflects the absence of marine-related maintenance activities, easier site accessibility, and reduced complexity of cleaning and repair operations. Both systems assume a 2.0\% annual O\&M escalation rate to account for inflation over the 25-year project lifetime.

Despite the higher investment cost, the FSPV system exhibited superior energy generation performance due to the natural cooling effect of water, which improves PV module operating efficiency and reduces thermal losses. As a result, the FSPV system generated approximately 113.24 million kWh during the first year of operation, achieving a capacity factor of 24.4\% and an energy yield of 2,137 kWh/kW. In comparison, the LBPV system generated 106.86 million kWh in Year 1, with a capacity factor of 21.4\% and an energy yield of 1,985 kWh/kW. This represents a 5.6\% higher annual energy generation for the FSPV system, translating to a 3.0 percentage point advantage in capacity factor. The performance ratio of the LBPV system was simulated at 0.78, while the FSPV system benefits from the cooling effect to achieve a higher effective performance ratio. This enhanced energy yield contributed to an increased annual revenue of approximately 24.01 million USD for the FSPV system, while the LBPV system generated approximately 22.65 million USD annually. \\
The LCOE analysis revealed that the FSPV configuration exhibited a higher LCOE of 0.0476 USD/kWh compared to 0.0403 USD/kWh for the LBPV system, meaning the floating system produces electricity at a cost approximately 18\% higher than the land-based alternative. This difference aligns with published literature, which typically reports FSPV LCOE values ranging from 18\% to 30\% higher than conventional LBPV systems. The higher LCOE of the FSPV system is primarily attributable to its elevated capital expenditure, despite its superior energy yield. However, both LCOE values remain substantially lower than the current industrial electricity tariff in Nicaragua (0.212 USD/kWh), indicating that both technologies can deliver significant cost savings compared to grid electricity purchase. The LCOE was calculated using a discounted cash flow methodology over the 25-year project lifetime, incorporating a 12\% discount rate, 0.5\% annual module degradation, and 2.0\% annual O\&M escalation.

A 70:30 debt-to-equity financing structure with an 8\% loan interest rate over 15 years was assumed for both configurations. The FSPV system achieved a higher NPV of 67.8 million USD compared to 59.6 million USD for the LBPV system, representing a 13.8\% higher total shareholder value over the project lifetime. This difference is primarily driven by the FSPV system's superior revenue generation (approximately 1.36 million USD more per year), which, when accumulated over 25 years and discounted at 12\%, outweighs its higher initial CAPEX and annual O\&M costs. The IRR analysis further supports the financial attractiveness of both systems, with the FSPV system achieving 24.1\% and the LBPV system achieving 22.3\%. Both values substantially exceed the 12\% discount rate (hurdle rate), indicating strong investment potential. The higher IRR of the FSPV system suggests that the additional capital invested in floating infrastructure generates attractive incremental returns.
The discounted payback period, which accounts for the time value of money at a 12\% discount rate, was calculated as 5.1 years for the FSPV system and 5.4 years for the LBPV system. The slightly faster payback of the FSPV system reflects its higher annual net cash flows, which allow the initial equity investment to be recovered more quickly despite the higher upfront capital requirement. For comparison, the simple payback periods (excluding debt service, taxes, and discounting) were 3.3 years for FSPV and 2.9 years for LBPV, demonstrating that both systems recover their initial capital relatively quickly on an undiscounted basis. The net cash flow to equity in Year 1 was calculated as 16.75 million USD for FSPV and 16.79 million USD for LBPV, indicating comparable immediate returns to investors despite the different investment levels.

Overall, the obtained results indicate that although FSPV systems involve higher initial investment costs (13.3\% higher CAPEX), their enhanced energy generation capability (5.6\% higher annual output) and improved long-term financial returns (13.8\% higher NPV, 1.8\% higher IRR) can offset the additional expenses. The floating system achieved a discounted payback period of 5.1 years compared to 5.4 years for the land-based alternative. The results suggest that floating solar technology offers a promising and economically attractive alternative to conventional land-based PV systems, particularly under conditions of limited land availability, existing water bodies, high ambient temperatures where cooling effects provide meaningful efficiency gains, and favorable regulatory environments with tax incentives.

\subsubsection{Economic analysis of the 398 MW FSPV system at El Cajón Dam, Honduras}
The economic feasibility of the proposed 398 MW floating solar photovoltaic (FSPV) system at the El Cajón hydroelectric reservoir in Cortés, Honduras, was evaluated based on the projected annual energy generation obtained from simulation results and the existing hydroelectric infrastructure available at the site. Compared to standalone floating solar installations such as the system analyzed for Lake Managua, the colocated hydro-floating configuration offers several economic advantages due to the availability of pre-existing substations, transmission lines, grid interconnection facilities, access roads, and operational infrastructure associated with the existing hydropower plant. These shared assets can substantially reduce the overall capital expenditure (CAPEX) and improve the long-term financial viability of the project.

For the Lake Managua FSPV system in Nicaragua, a CAPEX of 1,500 USD/kW was assumed, representing a typical standalone floating solar installation with no existing grid infrastructure at the site. This value is consistent with recent literature reporting standalone FSPV costs ranging from 1,200 to 1,800 USD/kW depending on site-specific conditions. However, for the El Cajón project, the CAPEX can be significantly reduced due to the colocation advantage. The existing hydropower plant at El Cajón already provides substation facilities, transmission lines to the national grid, access roads, and operational infrastructure, all of which would otherwise need to be constructed at substantial cost for a standalone project. By sharing these assets, the hydro-colocated FSPV system avoids approximately 30–40\% of the capital costs typically associated with standalone installations.

For the present study, a capital cost of approximately 950 USD/kW was therefore considered for the 398 MW FSPV system installed at the El Cajón reservoir. Based on this assumption, the total initial investment cost of the proposed system was estimated to be approximately 378.1 million USD. This represents a reduction of approximately 550 USD/kW (or 37\%) compared to the standalone Lake Managua system, directly attributable to the availability of existing hydropower infrastructure at the site \cite{chirwa2023floating, kakoulaki2023benefits}.

Among the total CAPEX, floating structures and pontoon systems constitute one of the major expenses in FSPV projects and generally account for 20–30\% of the total investment cost, similar to standalone systems, as these components are not shared with the hydropower plant. Therefore, the cost associated with floating platforms and floaters for the proposed El Cajón FSPV system was estimated to be approximately 75.6–113.4 million USD. Anchoring and mooring systems generally contribute an additional 5–10\% of the total CAPEX, corresponding to approximately 18.9–37.8 million USD for the proposed installation.
The annual operation and maintenance (O\&M) cost of FSPV systems is generally considered to be around 1\%–2\% of the total capital investment. For hydro-colocated systems, O\&M costs may be slightly lower due to shared operational resources with the existing hydropower plant, including site access, security, and maintenance personnel. In this work, the annual O\&M cost was assumed as 1.2\% of the total investment cost, corresponding to approximately 4.54 million USD per year. This is modestly lower than the 1.5\% assumed for the Lake Managua standalone system, reflecting the operational synergies with the existing hydro facility.

The El Cajón reservoir provides favorable conditions for large-scale FSPV deployment due to its extensive water surface area, high solar irradiance, and direct proximity to an established power generation and transmission network. In addition, the hybridization of floating solar with hydropower can improve grid stability, optimize reservoir utilization, reduce transmission losses, and enhance operational flexibility during periods of variable hydrological conditions. The cooling effect provided by the reservoir water can also improve photovoltaic module operating efficiency and reduce thermal losses, thereby increasing annual energy generation compared to conventional land-based PV systems operating under similar climatic conditions.

The annual energy generation of the proposed 398 MW FSPV system was obtained from simulation results. The system is expected to generate 821,744,960 kWh in the first year of operation, achieving a DC capacity factor of 23.6\%, an energy yield of 2,065 kWh/kW, and a performance ratio of 0.81.

The average electricity tariff considered for the economic analysis was based on the official industrial electricity rate in Honduras. The average electricity tariff is approximately 4.62 Honduran lempiras per kWh (HNL/kWh). Using an exchange rate of approximately 24.7 HNL per USD, this corresponds to approximately 0.187 USD/kWh. This estimation represents a realistic economic scenario and does not account for future increases in electricity prices, renewable energy incentives, or carbon credit benefits, which may further improve the financial attractiveness of the project.

Using the annual energy generation and the electricity tariff, the proposed FSPV system is expected to generate annual revenue of approximately 153.7 million USD during the first year of operation.
Under Honduran renewable energy regulations, the project is assumed to benefit from applicable tax incentives for renewable energy generation. A 70:30 debt-to-equity financing structure with an 8\% loan interest rate over 15 years was assumed for the financial analysis, consistent with the Lake Managua analysis. A discount rate of 12\% was applied to calculate the Net Present Value (NPV) and discounted payback period. A project lifetime of 25 years with an annual module degradation rate of 0.5
Based on the estimated annual revenue, annual O\&M expenses, and capital investment, the net annual cash flows of the FSPV system were determined for further financial assessment. Subsequently, the Levelized Cost of Energy (LCOE), Net Present Value (NPV), Internal Rate of Return (IRR), and discounted payback period were evaluated to determine the long-term economic viability of the proposed FSPV installation at El Cajón. Table \ref{tab:398MWFSPV} summarizes the key financial metrics for the proposed 398 MW FSPV system.

\begin{table}[t]
\small
\centering
\caption{398 MW FSPV system (El Cajón, Honduras) – Financial results}
\label{tab:398MWFSPV}
\renewcommand{\arraystretch}{1.2} 
\begin{tabular}{p{6cm} p{3cm} }
\hline
\textbf{Parameter} &
\textbf{Value} \\
\hline
Installed capacity (MW) &
398 \\

Year 1 AC energy generation (kWh) &
821,744,960 \\

Capacity factor (\%) &
23.6\%\\

Energy yield (kWh/kW) &
2,065 \\

Performance ratio &
0.81 \\

Total CAPEX (million USD) &
378.1 \\

CAPEX per kW (USD/kW) &
950 \\

Annual O\&M (million USD)	&
4.54 \\

Electricity tariff (USD/kWh) &	
0.187 \\

Annual revenue (Year 1) (million USD) &	
153.7 \\

LCOE (USD/kWh) &	
0.045 \\

NPV (million USD) &	
512 \\

IRR (\%) &	
22.1\\

Discounted payback period (years) &	
5.6\\

\hline
\end{tabular}
\end{table}

The calculated LCOE of 0.045 USD/kWh is highly competitive compared to both regional and global benchmarks. This value is lower than the LCOE of 0.056 USD/kWh reported for a 50 MW hydro-integrated FSPV system in Bangladesh and compares favorably with offshore floating solar projects in Southeast Asia, which achieve LCOE values below 0.06 USD/kWh in Thailand and Malaysia. The LCOE is also significantly lower than the 0.0476 USD/kWh calculated for the standalone Lake Managua FSPV system, demonstrating the economic benefit of colocation with existing hydropower infrastructure.
The NPV of approximately 512 million USD indicates strong positive shareholder value over the project lifetime. The IRR of 22.1\% substantially exceeds the 12\% discount rate, confirming that the project generates attractive returns for investors. The discounted payback period of 5.6 years indicates that the initial equity investment is recovered relatively quickly, which is favorable for project financing and risk assessment.
The strong financial performance of the proposed system is attributable to several factors specific to the El Cajón site: shared infrastructure avoiding new substation and transmission line construction (reducing CAPEX by approximately 37\% compared to standalone systems), high solar irradiance in the tropical location, the cooling effect of the reservoir enhancing PV efficiency, and economies of scale from the 398 MW capacity.

\section{Special Use Cases Under Different Application Scenarios}

\subsection{Colocation for AI data center plants - The case study of Guyana}
    
    Floating photovoltaic systems and water-adjacent data centers do create a mutually reinforcing industrial ecosystem. On the one hand, FSPV reduces land use, improves panel performance through passive water cooling, and can lower evaporation from reservoirs or cooling ponds \cite{bhattacharya2024energy}, whereas on the other hand, the data center benefits from a renewable local power supply. As grid interconnection delays lengthen and hyperscale demand rises, this colocated architecture offers a practical path toward self-supplied digital infrastructure built around renewable generation, storage, and grid-forming power electronics.

    \textit{Why the pairing works} - 
    Floating PV is attractive because water naturally cools the panels, which can improve output compared with land-based arrays, and it also avoids using valuable land that data centers often compete for. In addition, FPV can reduce evaporation from reservoirs or cooling ponds, which is especially useful in water-stressed regions where data centers also need reliable cooling resources. For a data center, that means the site can become both an energy node and a water-energy efficiency asset, rather than just a heavy load on the grid.
    
     \textit{Why this matters now} - 
    Grid congestion is now a binding constraint on data center growth, especially in Europe and the US, where connection times in legacy hubs average 7 to 10 years, and where rising demand is concentrating in already crowded markets. To put this in perspective, as of the end of 2024, there were 10,300 projects actively seeking grid interconnection in the U.S., representing 1,400 GW of generation and approximately 890 GW of storage \cite{rand2024queued}. In parallel, recent reporting shows the US regions, such as ERCOT, are facing very large data center-driven load growth, which is intensifying pressure on power systems and prices. Put plainly, many new compute campuses cannot wait for slow grid expansion, so they are moving toward self-supplied, behind-the-meter power systems.

    \textit{Primary drivers} - 
    \begin{itemize}
    \item The global bottleneck in grid access - In Europe, data center grid connections can take 7 to 10 years in legacy hubs, and grid congestion is already shaping (and limiting) where projects can be built \cite{nejla2025overcoming}.

    \item The behind-the-meter pivot is accelerating - Developers are increasingly considering onsite microgrids, batteries, and firming assets because the grid cannot always deliver power on the schedule hyperscale compute demands \cite{ghosh2026scalable, reinhardt2025ai}.

    \item Engineering stability is achievable without synchronous machines - Grid-forming converters, virtual inertia, and tuned control strategies are being developed to provide frequency support and stability \cite{abudyak2026mitigating} in low-inertia and inverter-dominated systems.

    \item Onsite hybrid generation - A renewable-heavy, battery-backed, grid-optional site can combine FSPV, storage, and firm capacity \cite{kwon2020ensuring} while preserving utility-grade reliability (Tier III data centers with 99.982\% and Tier IV data centers with 99.995\% availability) through layered redundancy and control.
    \end{itemize}

\begin{figure*}[!htbp]
    \centering
    \includegraphics[width=\textwidth]{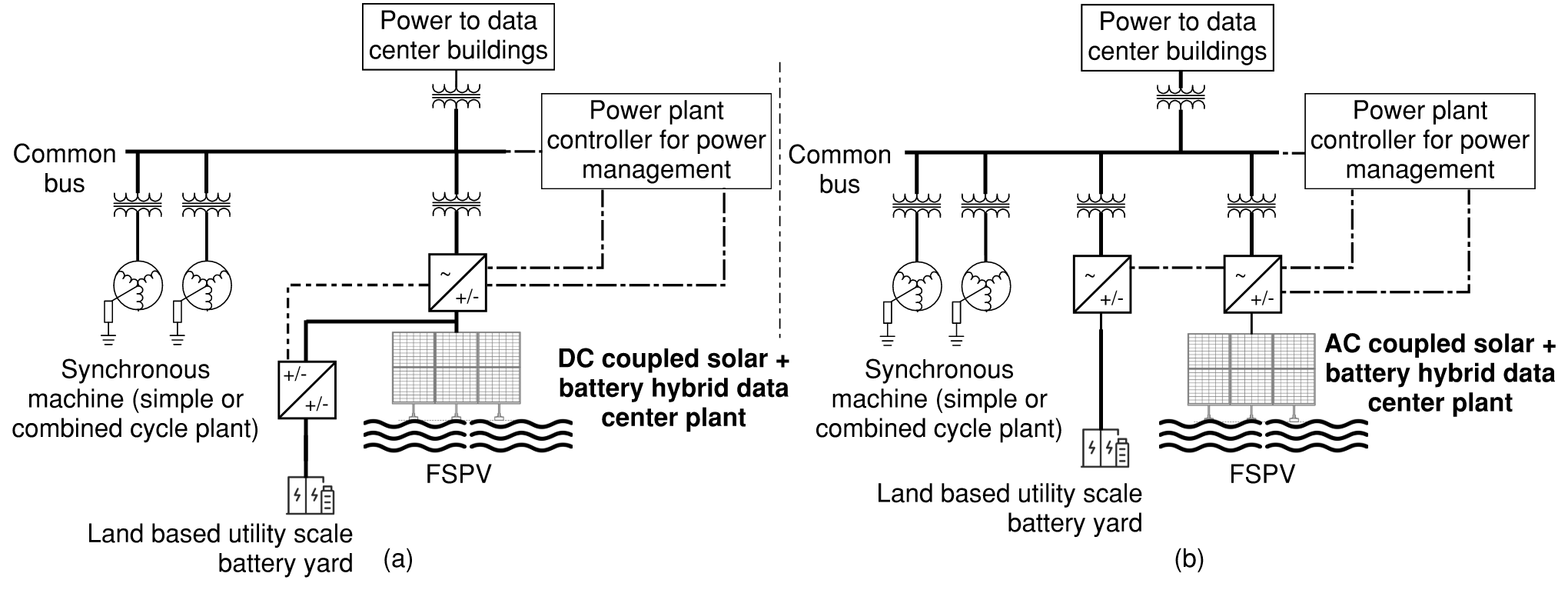}
    \caption{Hybrid plant architecture with synchronous machine and DC-coupled FSPV + battery power plants. (b) Hybrid plant architecture with synchronous machine and AC coupled FSPV + battery power plants.}
    \label{fig:Composite_Photo_Hybrid}
\end{figure*}

    For such off-grid hybrid co-located datacenters consisting of synchronous generators and FSPV systems with integrated land based battery energy storage power plants serving a data center cluster, two distinct power delivery architecture exist \cite{ghosh2026scalable}. As illustrated in Figure \ref{fig:Composite_Photo_Hybrid} (a) the DC coupling architecture is more efficient for new installations, while AC coupling, as in Figure \ref{fig:Composite_Photo_Hybrid} (b) may be better for retrofits on existing PV standalone plants. A detailed comparison of DC-coupled and AC-coupled solar PV systems integrated with battery energy storage systems (BESS) was performed by the authors in \cite{bitaraf2025powering}, and shows that DC-coupled configurations generally deliver better efficiency, greater reliability, and lower overall cost. They can also improve operational longevity by reducing conversion losses and simplifying the power pathway. In addition, the study demonstrated that DC-coupled architecture can lower capital expenditure and reduce the levelized cost of energy, with grid connection cost savings reaching as much as 20\%. Overall, DC-coupled systems present a strong option for independent power producers developing renewable energy portfolios to supply data centers, supporting both sustainability and energy security amid rapidly growing demand.

    Based on Table \ref{tab:fspv_potential} and Figure \ref{fig:waterbodiesofinterst}, Guyana contains three geographically clustered water bodies with strong potential for floating solar deployment: Lake Capoey, Lake Mainstay, and Lake Tapakuma. Taken together, these sites could support an estimated 95 MWac of FSPV capacity, with 203,708,624 kWh annual AC energy generation, and a performance ratio of 0.83, per estimations documented in Table \ref{tab:lake_Capoey_metrics}. Their close proximity to Queenstown also makes them particularly attractive because they can benefit from nearby electric transmission and distribution infrastructure, reducing the need for substantial new grid-connection works. As a result, a data center located in this area could draw on both the renewable electricity produced by the FSPV installations and the existing grid supply, improving energy reliability while supporting a lower-carbon power mix.
    
\subsection{Colocation within hydro reservoir area - The case study of FSPV over El Cajón dam in Cortés, Honduras}

\begin{figure*}[!htbp]
    \centering
    \includegraphics[width=\textwidth]{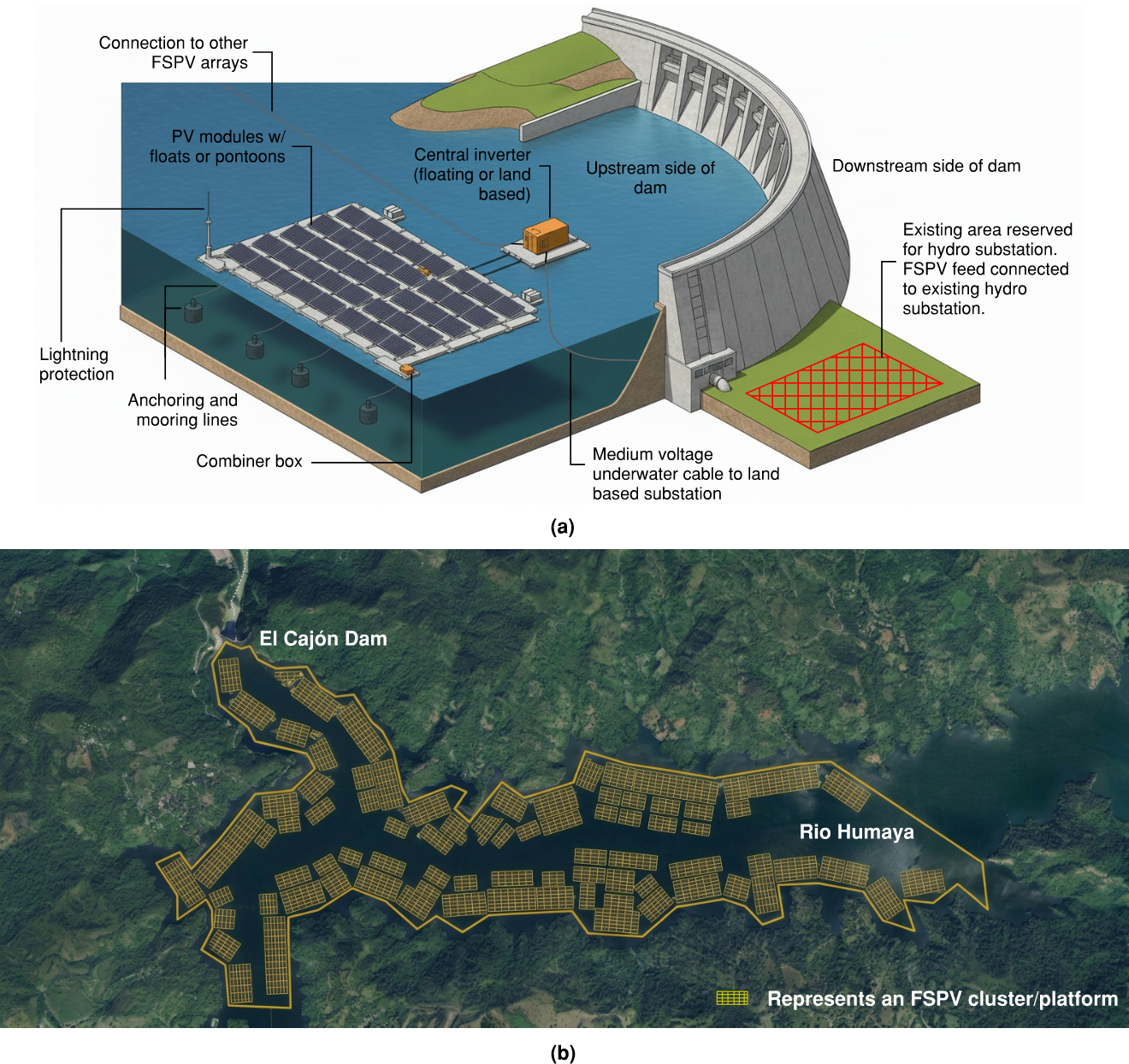}
    \caption{(a) Application proof of sketch of FSPV over El Cajón dam in Cortés, Honduras, (b) Proof of concept of a high concentration of FSPV colocated within the hydro-reservoir area of the El Cajón dam, Honduras. The area within the yellow polygon covers 1,030 acres of surface area.}
    \label{fig:hydrodam_El_Cajon_Dam_}
\end{figure*}

    When floating solar photovoltaic (FSPV) systems are colocated with existing hydroelectric reservoirs, they can leverage the pre-existing electrical infrastructure, including substations, transmission lines, and grid interconnection facilities that were originally developed for hydropower generation. This shared use of assets can significantly reduce capital expenditure (CAPEX) by avoiding the need for new land acquisition, civil works, and power line construction. A comparable study done by \cite{ghiasi2025strategic} with conceptual FSPV implementations in Iran's dam reservoirs estimated an average CAPEX of USD 700,000/MW and a competitive LCOE of USD 0.048 - 0.065/kWh. Furthermore, because these FSPV installations feed power directly into the established grid, they do not require independent microgrid arrangements. The renewable electricity generated by the FSPV system effectively offsets power that would otherwise be produced from fossil-based or biofuel generation, thereby lowering overall carbon intensity while strengthening the hybrid operation of the hydro-solar system.
    Figure \ref{fig:hydrodam_El_Cajon_Dam_} (a) illustrates the conceptual layout of such an FSPV system behind a reservoir dam, with Figure \ref{fig:hydrodam_El_Cajon_Dam_} (b) providing a satellite view of the FSPV cluster platforms covering approximately 1,030 acres of surface area across Rio Humaya near the El Cajón dam.
    
    To evaluate the role of FSPV as a cornerstone of Honduras’s transition toward a sustainable energy future, we examined greenhouse gas emissions by sector for the country in 2022, as summarized in Table \ref{tab:Honduras_GHG}. The power and transportation sectors emerged as the largest contributors to CO2, NOx, and SO2 emissions. Consistent with the scope of this manuscript, the analysis was concentrated on the power sector, which had a total installed generation capacity of 3,159.33 MW in 2022  \cite{IRENA}, of which 54.6\% or 1,725 MW was supplied by fossil and biofuel-based generation, as per Figure \ref{fig:elegensource}. Based on the projected 398 to 557 MWac generation capacity at the project site, as presented in Table \ref{tab:fspv_potential}, we estimated the potential reduction in CO2, NOx, and SO2 emissions under scenarios in which FSPV at El Cajon displaces 50\%, 25\%, and 10\% of fossil and biofuel generation, with results reported in Table \ref{tab:fspv_ghg_displacement}. From an environmental perspective, deployment of FSPV at the El Cajon dam could displace fossil fuel-based electricity generation and avoid the release of 128.8 tons of CO2, 2.2 tons of NOx, and 2.6 tons of SO2 per MW, underscoring the significant air quality benefits that commercial-scale FSPV can provide.

\begin{table}[t]
\small
\centering
\caption{Greenhouse gas emission by sector for Honduras in 2022.}
\label{tab:Honduras_GHG}
\renewcommand{\arraystretch}{1.2} 
\begin{tabular}{p{3cm} p{2.5cm} p{2.5cm} p{2cm}}
\hline
\textbf{Emission by sector $(2022) ^1$} &
\textbf{CO2 (tons)} &
\textbf{NOx (tons)} &
\textbf{SO2 (tons)} \\
\hline
Buildings &
622,473 &
3,770 &
4,509 \\

Fuel exploitation &
17,633 &
- &
- \\

Industrial combustion &
1,381,180 &
5,716 &
8,396 \\

Power industry $^2$ &
3,245,980 &
54,809 &
66,096 \\

Process industry &
668,869 &
- &
- \\

Transport &
4,601,960 &
36,074 &
7,837 \\

Water &
32,587 &
- &
- \\

\hline
\end{tabular}
\\[3pt]
\footnotesize\textsuperscript{$^1$}Source: \cite{EDGAR}.\\$^2$ Denotes sector of interest in evaluating potential tons of greenhouse gas reduced due to displacement of fossil and bio-fuel with FSPV, as computed in Table \ref{tab:fspv_ghg_displacement}.
\end{table}

\begin{table}[t]
\small
\centering
\caption{Computing tons of greenhouse gas reduced by FSPV deployment at El Cajón dam due to the displacement of fossil and bio-fuel with FSPV}
\label{tab:fspv_ghg_displacement}
\renewcommand{\arraystretch}{1.2} 
\begin{tabular}{p{4.5cm} p{3.5cm} p{3.5cm} p{3.5cm}}
\hline
\textbf{Greenhouse gas in consideration} &

\multicolumn{3}{l}{\textbf{Tons of GHG reduced due to displacement of fossil and bio-fuels with FSPV}} \\
\textbf{} &
50\% dispacement (199 - 279 MW)* &
25\% dispacement (99 - 139 MW)* &
10\% dispacement (39.8 - 56 MW)* \\
\hline
CO2 (tons) &
374,464 - 524,061 &
187,232 - 262,031 &
74,893 - 104,812 \\

NOx (tons) &
6,323 - 8,849 &
3,161 - 4,424 &
1,265 - 1,770\\

SO2 (tons) &
7,625 - 10,671 &
3,812 - 5,336 &
1,525 - 2,134 \\
\hline
Average tons of CO2 offset/MW of fossil and bio-fuel generation &
260.4 &
130.2 $^\dagger$ &
52.09 \\
\hline
Average tons of NOx offset/MW of fossil and bio-fuel generation &
4.4 &
2.2 $^\dagger$&
0.88 \\
\hline
Average tons of SO2 offset/MW of fossil and bio-fuel generation &
5.3 &
2.65 $^\dagger$&
1.1\\
\hline

\end{tabular}
\\[3pt]
\footnotesize *Based on 3159.33 MW of generation capacity for Honduras (2022) \cite{IRENA}, with 54.6\% (1725 MW) generation from fossil and bio fuels (54.6 \% based on Figure \ref{fig:elegensource}), and with an assumed potential for FSPV generation between 398 - 557 MW, based on data from Table \ref{tab:fspv_potential}. \\
$^\dagger$ indicates average tons of CO2, NOx, and SO2 that could be offset with FSPV implementation in Honduras at 25\% displacement of fossil and bio-fuels with FSPV. \\
See Data Availability Statement at the end of this manuscript for link to data.
\end{table}

\section{South American energy policies favoring renewable and low-carbon energy sources}
An understanding of successful and favorable renewable energy policies from South American countries that score a cumulative high score in terms of GDP per capita and electricity consumption per capita may help pick up on the linkage between the renewable energy policies promoting the low-carbon transition in these nations. Argentina, Brazil, and Chile were selected based on the aforementioned criteria, and the energy policies for these countries over the last twenty years were studied, and filtered to focus on those that have a positive outcome in advancing the nations' renewable energy goals; these policies are summarized in \S\ref{Argentina} - \S\ref{Chile}. 
 
\subsection{Argentina}
\label{Argentina}
Argentina’s renewable energy transition has been closely structured around successive policy frameworks that progressively linked resource promotion with explicit low carbon objectives. Early measures such as Law 26.190 in 2006 introduced quantitative targets for renewable electricity and fiscal incentives, establishing an initial 8 percent goal and signalling a shift away from purely fossil based expansion. These foundations were significantly deepened by Law 27.191 in 2015, which tightened and extended targets to 20 percent of electricity consumption by 2025 and enabled the RenovAr auction program and associated guarantee mechanisms to attract large scale private investment in wind and solar capacity. The emergence of markets such as MATER (Mercado a Término de Energías Renovables) for direct contracts between generators and large consumers further reinforced this policy-induced demand, progressively embedding non-hydro renewables as a structural component of the power mix.

On the transition side, these promotion instruments have coincided with a marked change in Argentina’s generation profile and climate policy narrative, even if formal targets have not always been met on schedule. Installed wind and solar capacity has grown rapidly since 2016 under RenovAr and subsequent schemes, contributing to record renewable penetration levels above 7 GW of non-hydro capacity and renewable shares of around one-fifth of electricity demand by the mid-2020s under the Law 27.191 framework. At the same time, the integration of energy sector measures into updated nationally determined contributions and long-term climate strategies has reframed renewable deployment as a central mitigation pillar rather than a purely diversification tool. Recent analyses highlight both the significant emissions reduction potential associated with continued scaling of wind and solar, and the need for complementary grid, flexibility and market reforms to sustain higher shares beyond the original 2025 target horizon. Table \ref{tab:govtpolicies_Argentina} provides a comprehensive overview of these enabling energy policies in Argentina.

\begin{table}[t]
\small
\centering
\caption{Key renewable energy and climate policies in Argentina enabling large‑scale adoption of low‑carbon energy sources.}
\label{tab:govtpolicies_Argentina}
\renewcommand{\arraystretch}{1.2} 
\begin{tabular}{p{3.5cm} p{3cm} p{2cm} p{5cm}}
\hline
\textbf{Policy / Instrument} &
\textbf{Category} &
\textbf{Year of enactment} &
\textbf{Main goals (renewables / low‑carbon focus)} \\
Argentina Renewable Energy Auction (RenovAR and RenovAR 2) &
Resolution/auctions &
2016 and 2017 &
RenovAr is Argentina's government tender program to increase renewable energy capacity, with RenovAr 2 being the second round of bidding designed to award additional wind, solar, and biomass projects. \cite{taretto2025exploring}\\

\hline
Law 27.424 &
Law &
2017 &
Create a legal framework for small‑scale distributed renewable generation by end‑users, enable net‑billing/compensation schemes, and support distributed resources to contribute to the national targets. \cite{taretto2025exploring} \\
\hline

Law 27.520 &
Framework climate law &
2019 &
Establish a national climate change policy framework. Establishes the minimum budgets for environmental protection to guarantee adequate actions. \cite{law27520} \\

\hline
National Climate Change Adaptation and Mitigation Plan (PNAMCC) &
National plan / strategy &
2019–2020 (approval and updates) &
Operationalize Law 27.520 by defining sectoral mitigation and adaptation measures, with power‑sector actions centered on scaling renewables and improving energy efficiency to meet 2030 emissions objectives. \\
\hline
National Plan for Mitigation and Adaptation to Climate Change &
International agreement implementation &
2022 &
Commit to economy‑wide mitigation targets (e.g., not exceeding 359 MtCO2e by 2030 and later strengthening ambition), thereby driving policies to expand renewables, improve efficiency, and align the power sector with a low‑carbon pathway. \cite{argentinalaw} \\

\hline

\end{tabular}
\end{table}

\subsection{Brazil}
\label{Brazil}
Brazil’s trajectory in linking renewable energy policy with its low‑carbon transition builds on an already high share of renewables in the power mix, historically dominated by large hydropower and sugarcane ethanol. Early diversification efforts responded to hydro‑dependency and oil price shocks, with instruments such as PROINFA and subsequent auction schemes stimulating wind, biomass and small hydro, while long‑standing ethanol and biodiesel mandates embedded biofuels into transport decarbonization. Over the 2000–2020 period, successive energy plans and regulatory adjustments sought to maintain a predominantly renewable electricity matrix, even as periodic droughts exposed hydrological risks and prompted targeted incentives for wind and, later, solar PV. By the early 2020s, Brazil’s electricity mix remained over 80 percent renewable, with rapid growth of wind and utility‑scale and distributed solar progressively reducing the relative weight of large hydro and broadening the portfolio of low‑carbon resources.

In parallel, Brazil’s climate policy architecture increasingly framed these sectoral instruments within an explicit long‑term decarbonization narrative, including commitments under its Nationally Determined Contributions and planning documents such as the Ten‑Year Energy Expansion Plans and the emerging net‑zero by 2050 agenda. Recent policy packages, such as RenovaBio for advanced biofuels and new legislation on “Fuel of the Future” and green hydrogen, aim to couple expanded renewable supply with deeper electrification and low‑carbon fuels in hard‑to‑abate sectors, while maintaining system reliability and affordability. Together, these measures signal a gradual shift from opportunistic use of abundant hydropower and biomass toward a more integrated energy‑transition strategy, in which diversified renewables, bioenergy, and emerging vectors like hydrogen support Brazil’s broader goal of a predominantly low‑carbon energy system by mid‑century. Table \ref{tab:govtpolicies_Brazil} summarizes these Brazilian energy policies.

\begin{table}[t]
\small
\centering
\caption{Key renewable energy and climate policies in Brazil enabling large‑scale adoption of low‑carbon energy sources.}
\label{tab:govtpolicies_Brazil}
\renewcommand{\arraystretch}{1.2} 
\begin{tabular}{p{3.5cm} p{3cm} p{2cm} p{5cm}}
\hline
\textbf{Policy / Instrument} &
\textbf{Category} &
\textbf{Year of enactment} &
\textbf{Main goals (renewables / low‑carbon focus)} \\
\hline
Law 9.427 – Creation of ANEEL and regulation of the electricity sector &
Law &
1996 &
Establish the national electricity regulator (ANEEL), organize the power sector, and create conditions for private investment, including in non‑large‑hydro renewable generation.
 \\

\hline
Law 10.438 &
Law / promotion program basis &
2002 &
Create the Incentive Program for Alternative Sources of Electricity (PROINFA) and the Energy Development Account (CDE) to expand wind, biomass, and small hydro, diversify the matrix, and support universal access. \cite{ruiz2010electricity}\\

\hline
Law 10.848 and Decree 5.163 – Power commercialization framework &
Law and decree &
2004 &
Law 10.848 of March 15, 2004, is the foundational legal framework for the modern Brazilian electricity sector, which established the new model for energy trading, auctions, and sector regulation. Decree 5.163 establishes guidelines for the commercialization of energy through auctions, with a focus on supply security and long-term contracts. \\

\hline
Law 13.576 – RenovaBio (National Biofuels Policy) &
Law / sectoral decarbonisation &
2017 &
Establish Brazil’s National Biofuels Policy (RenovaBio), set mandatory annual decarbonization targets for the fuel matrix, create certification of biofuels and tradable Decarbonization Credits (CBIOs) to reduce carbon intensity of transport fuels. \cite{brazilrenovlaw}\\

\hline
Law 12.187 update (PL 6.539/2019 approval) – PNMC alignment with Paris Agreement &
Law / update to climate framework &
2021 - 2022 &
Update the National Policy on Climate Change to align with the Paris Agreement, establish carbon‑neutrality by 2050 as a goal, and link NDC revision to scientific inventories, thus strengthening Brazil’s low‑carbon and renewable‑energy transition signal. \\
\hline

\end{tabular}
\end{table}

\subsection{Chile}
\label{Chile}
Chile’s renewable energy transition has been driven by a sequential tightening of regulatory obligations on electricity companies, which progressively reframed non‑conventional renewable energy from a niche option to a central pillar of the power mix. Law 20.257 first introduced a quota system that required new electricity supply contracts to source a rising share of delivered energy from non‑conventional renewable energy, starting at 5 percent and ramping toward 10 percent, which provided clear market signals and a predictable demand floor for wind, solar and other emerging technologies. Law 20.698, commonly known as the 20/25 Law, subsequently strengthened this framework by raising the obligation to 20 \% non‑conventional renewable energy by 2025, thereby catalyzing a rapid build‑out of utility‑scale wind and solar and positioning renewables as a least‑cost option in Chile’s competitive wholesale market. Chile’s Nationally Determined Contribution under the Paris Agreement then embedded these power‑sector dynamics within a broader low‑carbon strategy by linking rising renewable electricity shares, coal phase‑out milestones and absolute greenhouse gas emission limits to long‑term carbon‑neutrality objectives, thus consolidating earlier quota‑based promotion instruments into a coherent climate‑policy architecture that steers continued decarbonization of the electricity system. Table \ref{tab:govtpolicies_Chile} summarizes these enabling energy policies for Chile.

\begin{table}[t]
\small
\centering
\caption{Key renewable energy and climate policies in Chile enabling large‑scale adoption of low‑carbon energy sources.}
\label{tab:govtpolicies_Chile}
\renewcommand{\arraystretch}{1.2} 
\begin{tabular}{p{3.5cm} p{3cm} p{2cm} p{5cm}}
\hline
\textbf{Policy / Instrument} &
\textbf{Category} &
\textbf{Year of enactment} &
\textbf{Main goals (renewables / low‑carbon focus)} \\
\hline
Law 20.257 – Non‑Conventional Renewable Energies (NCRE Law) &
Law &
2008 &
Amend the General Law on Electrical Services to define non‑conventional renewable energy and mandate that a share of energy sold by generators with >200 MW capacity must come from NCRE, starting at 5\% of contracted energy (2010–2014) and rising annually to reach 10\% by 2024.
 \\

\hline
Law 20.698 – “20/25 Law” &
Law &
2013 &
Increase the NCRE obligation introduced by Law 20.257 by setting a target for 20\% of electricity generation to come from non‑conventional renewables by 2025, establishing a revised trajectory of annual percentage requirements for electricity producers. \cite{chile20698}\\

\hline
Chile’s Nationally Determined Contribution (NDC) under the Paris Agreement (updated NDC) &
International agreement implementation &
2015 (1st NDC), updated 2020 &
Establish economy‑wide mitigation targets and carbon‑budget‑type limits, including explicit reliance on rapid expansion of renewables, coal phase‑out, and energy efficiency to reach carbon neutrality by 2050.
\\
\hline
2014–2018 Energy Program &
Program incentive &
2014/2018 &
To achieve 45\% renewable energy share for new electrical capacity installed between 2014 and 2025. \cite{cabre2015renewable}
\\
\hline
\end{tabular}
\end{table}

\begin{figure*}[!htbp]
    \centering
    \includegraphics[width=\textwidth]{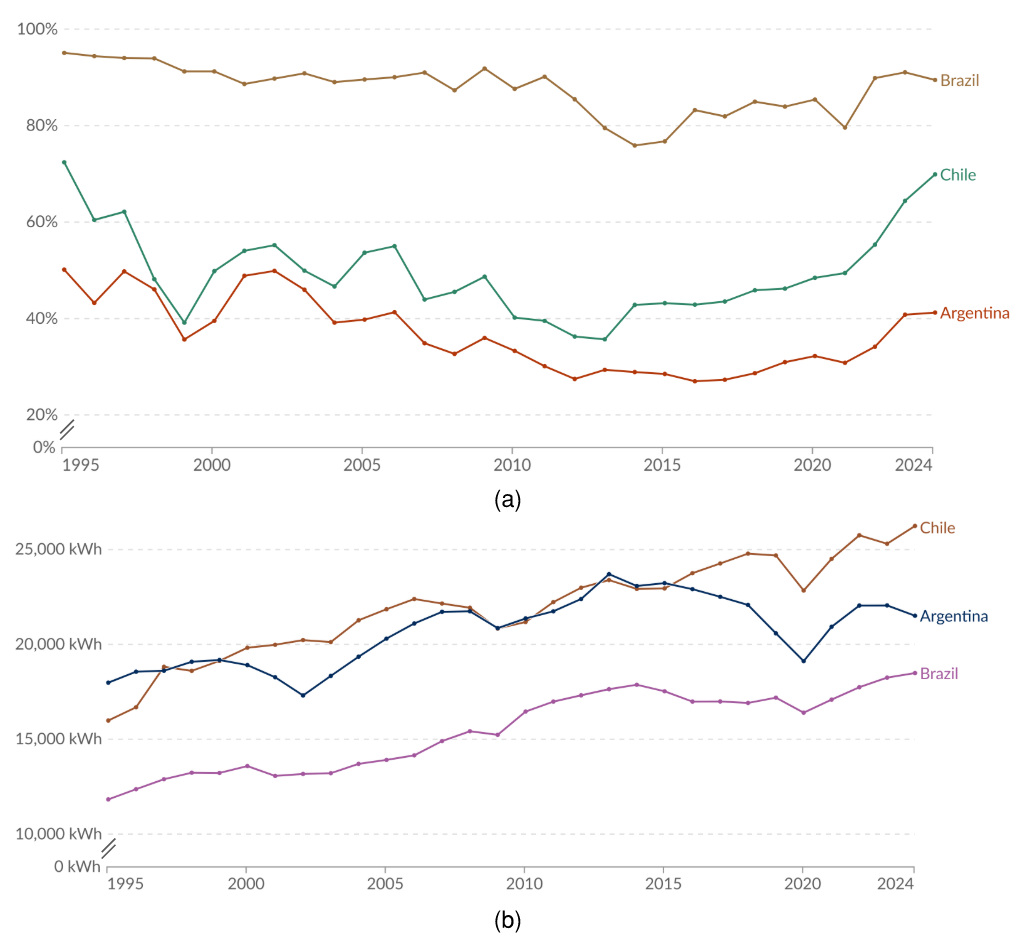}
    \caption{Analysis of cumulative variation of (a) share of electricity generated by low-carbon sources measured as a percentage of total electricity produced, and (b) energy use per person measured in kilowatt-hours per person. Data source: \cite{ourworldindata_1, ourworldindata_2} }
    \label{fig:Low_carbon_and_per_capita}
\end{figure*}

To gauge the effectiveness of these selected energy policies favoring renewable energy growth, the share of electricity generated by low-carbon sources
(measured as a percentage of total electricity produced) were tracked between 1995 - 2024, see Figure \ref{fig:Low_carbon_and_per_capita}(a) for Argentina, Brazil, and Chile. All three countries show an upward trend in low-carbon energy source utilization post-2010. During this time period, Brazil has been the forerunner and has constantly maintained its low-carbon energy source utilization at 75\% or higher. During the same period, Argentina boosted its low-carbon energy utilization by over 15\%, attributed primiarily due to its adoption of Law 27.424 favourable for small-scale distributed renewable energy resources, and the aggressive enactment of Law 27.520 establishing certain minimum standards for environmental protection regarding climate change adaptation and mitigation. In parallel, Chile has demonstrated an even superior growth with an approximate 40\% increase in its low-carbon energy utilization portfolio, primarily attributed to its 2023 “20/25 Law” and the country's adoption of Nationally Determined Contribution (NDC) under the Paris Agreement (updated NDC) in 2015. 

The correlation between these renewable energy and climate policies and the strong post-2010 growth of low-carbon energy sources is further established from the fact that all three countries were able to increase or sustain its energy use level per person. While Brazil and Chile showed an upward trend in the energy use per person (measured in kilowatt-hours per person), see Figure \ref{fig:Low_carbon_and_per_capita}(b), Argentina was able to maintain its energy use per person portfolio at an average of 23,000 kWh while increasing its low-carbon energy utilization by over 15\%. Energy policy makers in Nicaragua, Honduras, and Guyana are likely to benefit from a closer study of these energy policies from Argentina, Brazil, and Chile, when it comes to long-term adoption policies surrounding low-carbon and renewable sources, such as FSPV. 

\section{Conclusion, challenges, and future research directions}

\subsection{Conclusion}
This study establishes a comprehensive techno-socio-economic framework to evaluate the viability and systemic co-benefits of Floating Solar Photovoltaic (FSPV) systems within Latin America and the Caribbean, using location-specific case studies in Nicaragua, Honduras, and Guyana. While conventional land-based solar installations face extensive land-use constraints and thermal degradation in tropical environments, FSPV technology effectively circumvents land acquisition conflicts by utilizing existing underutilized water surfaces. The macro-level resource analysis confirms that Central and South American freshwater bodies possess the highest normalized floating solar generation potential globally, yielding up to 38.26 TWh per million acres.Rigorous microclimatic and technical simulations executed across capacities ranging from 50 MW to 398 MW demonstrate high operational efficiency and strong resource availability. The results show that the passive cooling effect of the water bodies successfully lowers module operating temperatures, yielding first-year energy generation values that translate into DC capacity factors above 20\% and exceptionally high annual energy yields exceeding 1,950 to 2,140 kWh/kW. In direct comparisons under identical regional weather profiles, FSPV systems consistently outperform conventional ground-mounted installations by 5\% to 6\% in total annual AC electricity output due to reduced thermal degradation and lower ground-level soiling losses.\\

From an economic and infrastructure perspective, the framework highlights the distinct advantages of standalone versus co-located configurations. Standalone installations, such as the 50 MW system on Lake Xolotlán in Nicaragua, display strong financial viability with a levelized cost of energy of 0.0476 USD/kWh, an internal rate of return of 24.1\%, and a discounted payback period of 5.1 years under local industrial tariffs. When scaled to hybrid hydro-FSPV configurations, the co-location advantages become even more compelling. At the 398 MW El Cajón reservoir site in Honduras, sharing pre-existing electrical substations, access roads, and grid interconnection assets reduces capital expenditures by approximately 37\% relative to standalone developments, lowering the capital cost to 950 USD/kW and the LCOE to an ultra-competitive 0.045 USD/kWh.\\

Furthermore, the environmental analysis reveals that large-scale FSPV deployment operates as a powerful instrument for regional decarbonization and water security. Hybrid hydro-solar operations protect vulnerable, hydro-dependent national grids against seasonal droughts by conserving water resources through surface evaporation reductions of up to 60\%. At a 25\% fossil-fuel grid displacement baseline, the 398 MW El Cajón development avoids the release of 130.2 tons of CO2, 2.2 tons of NOx, and 2.65 tons of SO2 per megawatt of thermal generation offset, offering massive air-quality and climate mitigation co-benefits. \\

Finally, the study establishes that innovative industrial architectures, such as DC-coupled FSPV-BESS microgrids, can successfully address severe grid interconnection bottlenecks and lengthy infrastructure delays for high-electricity-demand facilities like AI data center campuses. This is illustrated by the 95 MW coastal water body cluster in Guyana, which achieves a first-year annual AC energy generation of 203,708,624 kWh and an optimized performance ratio of 0.83, providing an independent, behind-the-meter green power pathway. By bridging these site-specific engineering findings with regulatory precedents from regional leaders, this framework provides a highly scalable blueprint to accelerate low-carbon investment, support local community economic development, and achieve universal clean energy access targets (SDG7) across emerging nations.

\subsection{Challenges and Mitigation Strategies} 

Despite the clear engineering and financial advantages, wide-scale commercialization of FSPV systems across Latin America faces distinct physical, structural, and social challenges:
\begin{itemize}
    \item \textit{Material degradation and structural stress-} Deploying electrical equipment over tropical water surfaces subjects materials to continuous high relative humidity, accelerated galvanic corrosion, and potential UV degradation. Large water bodies like Lake Cocibolca are also subject to severe wind-induced wave activity and hydrodynamic loading, which introduce mechanical stress on the mooring lines. To mitigate these issues, systems must mandate the use of marine-grade materials, high-density polyethylene structural floats with high typhoon-wind tolerances, salt-resistant panel coatings, and advanced virtual inertia controls in grid-connected power electronics.
    
    \item \textit{Ecological and limnological impacts-} Environmental concerns and mitigation strategies: The social and economic benefits people get from reservoirs could be affected by the installation of floating solar systems. For instance, if solar panels cover areas used for fishing, they might disrupt the livelihoods of local families who depend on the reservoir for food and income. In Lake Apanás, located in Jinotega, Nicaragua, many residents rely on small-scale fishing to support their households. Large floating solar arrays could make it harder to cast nets or move boats, and the shading they create might also change how fish and aquatic plants behave. In addition, floating solar panels could alter the lake’s natural appearance, which might reduce its appeal for tourism or recreation and even impact nearby property values. For these reasons, it’s important for social researchers and project developers to understand community concerns and find ways to make such renewable energy projects more acceptable to local people. 
    
    \item \textit{Social and socioeconomic conflicts-} Large floating solar arrays can physically disrupt pre-existing community assets, such as small-scale fishing zones, domestic water usage, tourism, and local property values. For example, communities surrounding Lake Apanás in Jinotega, Nicaragua, rely extensively on subsistence net casting, which could be restricted by anchored structural platforms. Developers must deploy participatory planning frameworks, execute community-led site mapping, and align project bounds to avoid critical aquaculture and recreational zones.
\end{itemize}

\subsection {Future Research Directions}

To further expand the boundaries of floating solar technology and optimize its long-term integration into regional power grids, future research should prioritize the following areas:

\begin{itemize}
    \item \textit{Advanced multi-renewable hybrid systems-} Investigating the operational dynamics, seasonal complementary profiles, and transient control frameworks of co-located FSPV, floating wind, and pumped-hydro storage installations to achieve stable, baseload-grade renewable power profiles.
    
    \item \textit{Next-generation marine materials-} Developing and field-testing advanced, bio-friendly anti-fouling coatings and polymer composites engineered to prolong mooring line and structural pontoon lifecycles under intense tropical moisture and salinity regimes.
    
    \item \textit{AI-enabled diagnostics and predictive modeling-} Deploying machine learning algorithms and real-time sensor streams to track structural anchoring tension, automate module soiling classification, and execute predictive maintenance tasks on submarine cabling networks.
    \item \textit{Empirical long-term tracking systems-} Conducting multi-year field tracking of water chemistry, regional biological indicators, and microclimatic parameters around operational utility-scale installations to establish data-driven regulatory baselines for tropical limnology.
    
\end{itemize}

\section*{Funding}
This research received no external funding. This work was independently conducted and self-funded by the authors.

\section*{Declaration}
The authors declare that they have no known competing financial interests that could have appeared to influence the work reported in this paper. The views, thoughts, opinions, and conclusions made in this material are solely those of the authors and don’t necessarily reflect the views of the authors' employer, organization, committee, or other group or individual. 

\section*{Data Availability Statement}
The data used and/or derived for this work can be accessed through the project's GitHub page at: \href{https://github.com/sghosh27/Floating-Solar-Photovoltaics-Pushing-the-Frontiers}{https://github.com/sghosh27/Floating-Solar-Photovoltaics-Pushing-the-Frontiers}

\section*{Generative AI Usage Statement}
During the preparation of this work, the author(s) used Claude (Anthropic) to assist with sentence formatting and grammar checking. After using this tool, the author(s) reviewed and edited the content as needed and took full responsibility for the content of the published article.



\inConf{
\bibliographystyle{IEEEtran}
\bibliography{references}
}
\inArxiv{
\bibliographystyle{tmlr}
\bibliography{references}
}


\end{document}